\documentclass[12pt,a4paper]{article}
\pdfoutput=1
\usepackage{jheppub}  
\usepackage{pdfpages}
\usepackage{verbatim}
\usepackage{amsmath,amssymb,amsfonts,cases,amsthm}
\usepackage{graphicx,caption,subcaption}
\usepackage{mathtools}  
\usepackage{dsfont}
\usepackage{varwidth}
\usepackage{slashed}
\usepackage[shortlabels]{enumitem}
\usepackage{physics}
\usepackage{hyperref}
\usepackage{textcomp}
\usepackage{simpler-wick}
\usepackage{nicefrac}
\usepackage{subcaption}
\usepackage{dynkin-diagrams}
\usepackage[export]{adjustbox}

\usepackage{tikz}
\usetikzlibrary{arrows,shapes,positioning,decorations.markings,calc}
\usepackage[tikz]{mdframed}
\tikzstyle arrowstyle=[scale=1]
\tikzstyle directed=[postaction={decorate,decoration={markings,
    mark=at position .55 with {\arrow[arrowstyle]{stealth}}}}]
\usepackage{ifthen}
\usepackage{multirow}
\usepackage{longtable}
\usepackage[vcentermath]{youngtab}
\usepackage{afterpage}
\usepackage{lscape}
\usepackage{rotating}
\usepackage{float}

\mdfdefinestyle{shortnote}{outerlinewidth = 0.2 , roundcorner = 2pt , leftmargin = 4 , rightmargin = 4 , backgroundcolor = gray!15 , linecolor = gray!25 , innertopmargin = 10pt , innerbottommargin = 10pt , everyline = true , splittopskip = 18pt , splitbottomskip = 10pt}

\newcommand{\D}{\mathrm{d}}
\newcommand{\ridoff}[1]{}

\newcommand{\Bal}{\begin{equation}\begin{aligned}}
\newcommand{\Eal}{\end{aligned}\end{equation}}

\title{A Walk Through \texorpdfstring{$Spin(1,d+1)$}{Spin(1,d+1)} \\}
\author{Vladimir Schaub}
\emailAdd{vladimir.schaub@kcl.ac.uk}
\affiliation{Department of Mathematics, King's College London, Strand, London, WC2R 2LS, UK}

\abstract{We construct unitary irreducible representation of the de Sitter group, that forms the basis for the study of $dS_{d+1}$ QFT. Using the intertwining kernel analysis, we discuss bosonic symmetric tensor, and fermionic higher-spinors. Particular care is given to the structure and construction of exceptional series and fermionic operators. We discuss the discrete series, and explain how and why the exceptional series give rise to seemingly non-invariant correlators in de Sitter. Using tools from Clifford analysis, we show that for $d>3$, there are no exceptional fermionic representations, and so no unitary (higher)-gravitino fields. We also consider the structure of representations for $d=3$ and $d=2$, as relevant for the study of $dS_4$, which is the only space allowing for unitary gravitino and its generalisation, and of $dS_3$.}

\begin{document}
\maketitle

\section{Introduction}

What we can say about a physical system revolves around its symmetries. Much of what we know of quantum field theories springs out from Wigner's program, and its identification between unitary irreducible representation and particles \cite{Wigner:1939cj,Weinberg:1995mt}. As it has become apparent from observation that our space-time is expanding \cite{Hubble:1929ig}, the leading incorporation of gravity to QFT at large scale is that of a background rigid de Sitter spacetime structure. The study of de Sitter QFT is a vibrant area of research, which touches on experiments, trying to understand the physics of the CMB \cite{Bousso:2007gp,Baumann:2020ksv,Baumann:2020dch,Baumann:2019oyu,Baumann2018}, and grapples with fundamental questions aiming to understand its physical properties at the level of correlation functions \cite{Bros:1990cu,Weinberg2005,Bros:2010rku,Sleight:2019mgd,Sleight:2019hfp,Gorbenko:2019rza,Sengor:2019mbz,Sleight:2020obc,Sengor:2021zlc,Sleight:2021iix,DiPietro:2021sjt,Hogervorst:2021uvp,Goodhew:2021oqg,Melville:2021lst,Heckelbacher:2022hbq,Schaub:2023scu,DiPietro:2023inn,Anninos:2024fty,Loparco:2023rug,Loparco:2023akg,Galante:2023uyf,Loparco:2024ibp}, of the wavefunction \cite{Anninos:2014lwa,Goodhew:2020hob,Salcedo:2022aal,Anninos2021}, or in a holographic perspective \cite{Strominger:2001pn,Deser:2003gw,Anninos:2011ui,Anninos:2017eib,Hertog:2019uhy,Pethybridge:2024qci} . 

It seems crucial then, to have a good physical understanding of the Hilbert space structure consistent with the de Sitter isometries, meaning representation theory of $SO(1,d+1)$, the conformal group, and its double cover, which is the one actually relevant for a quantum Hilbert space. The present work is a step in that direction. This is very clearly not the first work on the de Sitter group : the study of $SO(1,d+1)$ is a central element of the theory of non-compact Lie groups, with (non-exhaustive) references dating back more almost 80 years and spanning decades \cite{Thomas:1941aa,Bargmann:1946me,Gelfand:1947aa,Harish-Chandra:1947aa,Newton:1950aa,Dixmier:1961aa,Hirai:1962aa,Takahashi:1963aa,Ottoson:1968aa,Gavrilik:1975aa}. From a physicist perspective however, the most crucial early reference seems to be \cite{Dobrev:1977qv}, since it places the construction of the unitary representations in a footing that is close to how quantum field are manipulated ;  working with correlation functions, quotienting out null states, and making harmonic decompositions. This account, though extremely complete, only investigated a specific class of representation, the totally symmetric traceless tensors. A modern treatment along these lines is given in \cite{Sun:2021thf}.

This creates a problem, since the existing mathematical literature, most notably the very exhaustive \cite{Hirai:1962aa,Ottoson:1968aa,Gavrilik:1975aa}, are not formulated in a language that is anywhere close to how we think of building fields, using instead Gelfand-Tsetlin patterns and recursively building unitary generators. These tools are of course extremely powerful and interesting in and of themselves, but they hide a lot of the physical content beyond sophistication. So, even though they have been used successfully to discuss more general representation like antisymmetric tensors etc. \cite{Basile:2017aa}, this method is not transparent and an alternative, following the lines of \cite{Dobrev:1977qv}, is ultimately preferable. 

An apt illustration of the present situation, and the motivation for this work, is the case of the (partially)-conserved (massless in the bulk) tensor-spinors \cite{Deser:2001us,Bonifacio:2023prb,Bonifacio:2018zex,Bonifacio:2021mrf,Blauvelt:2022wwa}. These are not unitary irreducible representation in $dS_{d+1}$, except in $d=3$. This specific dimension dependent feature seems to have been mostly overlooked in the literature, hidden away in the list of indices and weights, until it was pointed out by \cite{Letsios:2023qzq,Letsios:2023awz,Letsios:2024nmf}, which rediscovered it by studying the mode functions of a few of these (partially)-massless higher-spinors across dimension.\footnote{The method used in \cite{Letsios:2023qzq,Letsios:2023awz,Letsios:2024nmf} does not construct the representation from scratch at the level of the Hilbert space itself, but rather shows that the explicit field mode solutions transform in a unitary representation.} Since this family of representation includes the gravitino and our space is well described at large distance by $dS_4$, this rather exceptional fact must be thoroughly understood, especially given supersymmetry's difficult cohabitation with expanding space \cite{Pilch:1984aw,Deser:2003gw,Anous:2014lia,Bergshoeff:2015tra,Hertog:2019uhy,Anninos:2023exn}. Beyond these exceptional representations, more mundane questions such as why there are no complementary series spinors do not seem to have received a simple physical explanation so far.

\begin{figure}[H]
	\centering
	\includegraphics[width = 0.45\textwidth]{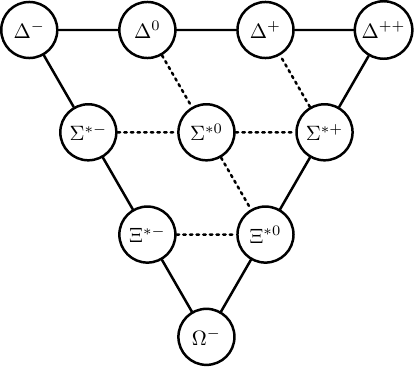}
	\caption{Some baryon with spin $3/2$ generated by QCD, organised in a decuplet \cite{Gell-Mann:1961omu}. Dressed multi-quark states are expected to overlap with excitations of all spin and in all families of unitary representations. }
	\label{fig:decuplet}
\end{figure}

The goal of the present work is to give a self-contained treatment, along the lines of \cite{Dobrev:1976vr,Sun:2021thf}, of the symmetric tensor-spinors unitary irreducible representations, leveraging the modern framework of conformal field theories, drawing notably \cite{Kravchuk:2016qvl,Karateev:2018oml,Karateev:2018uk,Iliesiu:2015qra,Iliesiu:2017nrv}. This might make the endeavour sound more restrictive than it really is. Precisely because supersymmetry is hard to come by in de Sitter, one cannot avoid studying the fermions by themselves and for their own sakes, even more so since fermions are plentiful in our universe. While spin-$1/2$ fermions have received some attention in the literature, see for example \cite{Pethybridge:2021rwf,Schaub:2023scu}, it would be wrong to discard higher-spinors. After all, there are plenty of known resonances, see fig. \ref{fig:decuplet}, which are of this type in our universe.

Because of the nature of these puzzling spinorial representations, which are in the exceptional (and actually discrete) series, one must confront the associated intricacies of these families of representations. On of the goal is to give here an exposition that makes the structure and construction of these representations clear, using the bosonic case as a review. Although essentially contained in the analysis of \cite{Dobrev:1977qv,Sun:2021thf}, this is still worthwhile from the standpoint of pedagogy and self-containment. This is not the only draw, since we tried to give a more detailed account of the peculiar, logarithmic yet conformally invariant, inner product of these representations and their quotient space structure, and relating it to the recent progress on the study of the discrete series in $dS_2$ and $dS_4$ \cite{Anninos:2023lin,Sengor:2023buj,RiosFukelman:2023mgq,Anninos:2024fty}. An important take away from our discussion, summarised in the appendix \ref{app:discrete}, is that the interesting features of these special representations is that they are \textit{exceptional}, not whether they are \textit{discrete}, since their features persist whether or not the group element have finite square-average.

Studying spinorial representations is an ambitious project by itself, and so this work will not address all of the associated issues. Notably, we will not consider the question of the characters of the representations, which can be understood in principle by applying the results of \cite{Hirai:1965aa,Sun:2021thf}, nor discuss the tensor-product decomposition \cite{Dobrev:1976vr,Penedones:2023uqc}. We leave these matters to future work. We will also not tackle the case $d=1$, relevant for the study of $dS_2$, simply because it was felt to be adequately studied in the present literature, see e.g. \cite{Kitaev:2017hnr,Sun:2021thf,Kravchuk:2021akc}.

The paper proceeds as follows. In sec. \ref{sec:boson}, we review the construction of unitary irreducible representations, using the symmetric traceleless tensor representations as an explicit case study. We explain the general strategy, detail the harmonic decomposition of the inner product. We investigate closely the exceptional points, first discussing scalars and then spinning representations. In sec. \ref{sec:spinor}, we perform the analogous analysis for tensor-spinor representations. The technique for the harmonic decomposition is not standard, and so we carefully derive them. The output is a proof that for $d>3$, there are no spinorial representations beyond the principal series because of a pairing of eigenvalues. What differentiates lower-dimension is clear : the Clifford algebra contracts, and one must study it case by case. This is done in sec. \ref{sec:dim}, for $d=3$ and $d=2$, where one can treat bosonic and fermionic representations at once. It is found that for $d=3$ only, at the exceptional points one can define a positive inner product, and so there are indeed some unitary partially-massless higher-spinors. We conclude with a discussion of the results and future research directions.

\section{Story of Bose :  Symmetric Tensor Representations}\label{sec:boson}

The goal of this section is to review the construction of the unitary irreducible representations (UIR) for the symmetric traceless tensors (STT). The central tool is Knapp-Stein intertwining kernel \cite{Knapp:1971aa}. Induced representations generate ``elementary representations", on which one tries to find an invariant, unitary inner product. The $L^{2}$ inner product, defines the Principal Series $\mathcal{P}_{\Delta}$. Another non-trivial inner product exists, which is a conformal 2pt. function. Through a partial wave analysis, we find bounds on the allowed ranges of $\Delta$, defining the complementary series $\mathcal{C}_{\Delta}$. For exceptional points, this analysis breakdowns, giving rise to a peculiar inner-product and the associated exceptional (sometimes discrete) series, $\mathcal{V}_{n}$, $\mathcal{U}_{\ell,t}$. We give a self-contained, conceptual discussion and try to give interpretation for this last procedure, which explains some features of fields in  de Sitter.

Our exposition is geared for the two other cases later considered ; fermionic representations, and lower-dimensions. Hence, care is given to the crucial role played by the harmonic decomposition to investigate the positive definiteness, and to the methodology for the investigation of exceptional points.

\subsection{Bruhat Decomposition And Intertwining Kernel}

The method of induced representations for semi-simple non-compact group follows by using a particular factorisation into subgroups \cite{Wigner:1939cj,Dobrev:1977qv,Knapp:1982aa,Knapp:1986aa} To consider unitary irreducible representations (UIR) of the group $G=SO(1,d+1)$, we start from the Bruhat decomposition into factors
\begin{align}
	N&=\left\{e^{b\cdot K}\right\} ,& A&=\left\{e^{\lambda D}\right\} ,& M &=\left\{e^{\frac{1}{2}\omega\cdot M}\right\} ,& 	 \widetilde{N}&=\left\{e^{x\cdot P}\right\} \, .
\end{align}
With $K$ the generator of special conformal transformations, $D$ of dilation, $M$ of rotations, and $P$ of translations. A representation of the subgroup $NAM$ lifts to a representation of the whole group. More precisely, starting from representations in the kernel of the nilpotent $N$, transforming as irreducible representations $(\rho,\Delta)$ of $M$ and $A$, one can act with an element of $\widetilde{N}$ to generate a full ``elementary'' representation; modulo some conditions regarding smoothness. The associated states can be associated to point in the homogeneous space $G/NAM \sim \mathbb{R}^{d}$, which  define a group action.\footnote{This construction is called the ``non-compact'' picture. It renders the $M$ subgroup evident. The ``compact'' or $K$-picture, makes the maximal compact subgroup $K=SO(d+1)$ manifest, and can be obtained from an inverse-stereographic projection. It is associated with the Iwasawa decomposition, whereby one writes a group element as $KAN$.}

We start with bosonic representation of the subgroup $SO(d)$, and we limit ourselves to fields in a representation $\rho$, with associated abstract indices $\alpha,\beta, \ldots$, given by a young tableau with a single row of $\ell$ boxes, i.e. symmetric traceless tensors of spin-$\ell$.\footnote{In the classic work \cite{Dobrev:1977qv}, these are called type I representations.} We use anti-hermitian generators, $Q^{\dagger}=-Q$. This defines complex conjugation and gives a framework to compute the norm of states, defining a notion of unitarity. The transformation properties of the elementary representation is conveniently encoded in braket notation\cite{Hogervorst:2021uvp,Sun:2021thf}, by defining a primary state
\begin{equation}
\begin{aligned}
	D\ket{\Delta,0}_{\alpha}&=\Delta\ket{\Delta,0}_{\alpha} \, , \\
	K_i\ket{\Delta,0}_{\alpha}&=0 \, ,\\
	M_{kl}\ket{\Delta,0}_{\alpha}&=-(\Sigma_{kl})_{\alpha}{}^{\beta}\ket{\Delta,0}_{\beta} \, ,
\end{aligned}
\end{equation}
where the matrices $\Sigma_{kl}$ are the matrix representation of the $M_{kl}$ in the representation $\rho$ of $SO(d)$, i.e. $D_{\rho}(M_{kl})=\Sigma_{kl}$. This state is irreducible under $A \sim D$, $N \sim K$, and $M$. The leftover $\widetilde{N}\sim P$ are translations that produce the whole conformal family  
\begin{align}
	\ket{\Delta,x}_{\alpha}&=e^{x\cdot P}\ket{\Delta,0}_{\alpha} \, ,
\end{align}
Which span the whole elementary representation. These states for finite $x$ transform as one would expect a primary with scaling dimension $\Delta$ to. We define the elementary representation as the functional space $\mathcal{F}_{\Delta,\ell}$, generically not unitary nor irreducible, of states created by wavefunctions of the form
\begin{align}
	\ket{\psi} = \int \D^{d}x \ \psi^{\alpha}(x)\ket{\overline{\Delta},x}_{\alpha} \, .
\end{align}
where we defined $\overline{\Delta}=d-\Delta$. The transformation properties of the states induces a transformation of the wavefunctions $\psi$, which is that of a primary operator with scaling dimension $\Delta$ in an Euclidean CFT$_d$, transforming under the conjugate representation $\rho_C$ of $SO(d)$. We now would like to stress the following important fact :  the base-states are a useful tool to induce the transformation properties of the wavefunctions $\psi$, which should ultimately be promoted to distribution-valued operators. All along however, they will be treated as functions to define the representations. It is the properly smeared states $\ket{\psi}$ which are part of the elementary representations, not the underlying $\ket{\overline{\Delta},x}$. Although in most instances one can work with ultra-local wavefunctions, this fails to capture some of the physics in the degenerate instances we will encounter, where one needs to quotient out null-subspaces. This distinction is crucial to understand the peculiar nature of the exceptional points later on.

We consider wavefunctions which are smooth and on which the generator of special conformal transformation is well behaved.\footnote{This can be understood as an artefact of the non-compact picture. In the compact picture, this simply translates into smoothness over $K$, the compactification of $\mathbb{R}^{d}$.} This implies an asymptotic condition, which is the mapping of the local continuity unto infinity,
\begin{align}
	\psi(x)\overset{x\to \infty}{\approx} \frac{1}{(x^2)^{\Delta}}\sum_{n} h_{n}\left(\frac{x^{\mu}}{x\cdot x}\right) \quad \mid \quad \tilde{x}\cdot \partial h_{n}(\tilde{x}) = n \,  h_{n}(\tilde{x}) \, ,
\end{align}
where $h_{n}$ is an homogeneous tensor-polynomial of order $n$, with scaling dimension $-n$ in $x$. This fixes the functional space $\mathcal{F}_{\Delta,\ell}$, and the elementary representation. 

It is a general result known as subrepresentation theorem \cite{Knapp:1986aa,Casselman:1989aa}, that any (unitary) irreducible representation can be found as an invariant subspace or quotient of elementary representation. Our goal is then to study the norm of states to find these spaces. It turns out that, except for isolated points, the representations are irreducible, making the analysis more streamlined.

To study unitarity and irreducibility, we need to define an inner-product $\bra{\cdot}\ket{\cdot}$ on $\mathcal{F}_{\Delta,\ell}$, which has to to satisfy multiple desiderata : positive definiteness, normalisability and irreducibility. It must take the generic form
\begin{align}
	\bra{\psi_1}\ket{\psi_2}& =\int \D^{d}x \ \D^{d}y \ \psi_1{}^{\dagger}_{\alpha}(x)G_{\overline{\Delta},\ell}(x,y)^\alpha{}_{\beta} \psi_2^{\beta}(y) = \big(\psi_1,G_{\overline{\Delta},\ell}[\psi_2]\big) \, .
\end{align}
The operation $(\cdot,\cdot)$ is the $L^2$ pairing of states transforming in conjugate representations.\footnote{Conjugation acts on representations of $A$, dilations, by sending $\Delta$ to $\overline{\Delta}=d-\Delta$} The function $G_{\Delta}$ is called the intertwining kernel, as convoluting a state of $\mathcal{F}_{d-\Delta,\ell}$ with $G_{\Delta,\ell}$ gives (barring exceptional cases) a state in $\mathcal{F}_{\Delta,\ell}$.\footnote{The operator $G_{\Delta}$ is also known as the Knapp-Stein intertwining operator \cite{Knapp:1971aa}, the intertwiner, and the shadow transform \cite{Ferrara:1972kab,Dolan:2004up,Simmons-Duffin:2014wb,Karateev:2018oml}.} 
\begin{equation}
	\begin{aligned}
	G_{\Delta,\ell}: \quad \mathcal{F}_{d-\Delta,\ell} &\to \mathcal{F}_{\Delta,\ell}  \, ,\\ \psi^{\alpha}(x)&\mapsto  \int   G_{\Delta,\ell}(x,y)^\alpha{}_{\beta} \psi^{\beta}(y) \D^{d}y \, .	
	\end{aligned}
\end{equation}

Conformal invariance implies Ward identities on the kernel, that in turn implies that $G$ transforms like a 2-point function of conformal primary fields with scaling dimension $\Delta$. Intuitively, this follows from $G_{\overline{\Delta},\ell}(x,y)=\bra{\overline{\Delta},x}\ket{\overline{\Delta},y}$.\footnote{This convention for the kernel differs from that of \cite{Sun:2021thf}, in which $G_{\Delta} = \bra{\overline{\Delta},x}\ket{\overline{\Delta},y}$.} This implies two qualitatively different cases : 

\begin{enumerate}[1.]
	\item $\Im(\Delta) \neq 0$. Conformal invariance imposes $\Delta + \Delta^{\dagger} = d$, giving $\Delta = \frac{d}{2}+i\nu$. The only allowed kernel is ultra-local, $G_{\frac{d}{2}-i\nu}(x,y)=\delta^{d}(x-y)$ and the fields transform in the irreducible\footnote{The intersection of the principal series with the real axis, $\nu=0$, asks for more care. For some choices of $\rho$, this point can belong to a reducible intertwiner, and so to a reducible elementary representation. This does not arise for STT representations, but it happens for spinors. This reduction simply amounts to quotienting by the null descendants of the differential equation satisfied by the state. This reducibility is not of an interesting type in general.} and unitary Principal Series $\mathcal{P}_{\frac{d}{2}+i\nu,\ell}$, which is equipped with the usual $L^2$ inner product $(\cdot,\cdot)$.
	
	\item $\Delta \in \mathbb{R}$, where the kernel is a non-trivial function of the two separation, and requires further analysis both for unitarity and irreducibility.
	\begin{enumerate}[i.]
	\item For almost all values of $\Delta$, the representation will be irreducible. For some range of $\Delta$, it will moreover be positive, giving rise to the complementary series $\mathcal{C}_{\Delta}$
	\item For some values of $\Delta$, the representation will be reducible, giving the exceptional (and discrete) series.
	\end{enumerate}
\end{enumerate}
The second case is the more involved one, to which we now specialise. To study the kernel, we use the index-free notation \cite{Costa:2011wa},  whereby we contract all free tensorial indices with null-vectors $z_i$,
\begin{equation}
	\begin{aligned}
	G_{\Delta,\ell}(x,y;z_1,z_2) &= z_1^{\mu_1}\ldots z_1^{\mu_\ell} G_{{\Delta},\ell}(x,y){}_{\mu_1\ldots \mu_\ell}{}^{\nu_1 \ldots \nu_\ell} z_{2,\nu_1}\ldots z_{2,\nu_\ell}  \\ 
	&=C_{\Delta,\ell}\frac{\left(z_1\cdot z_2 - 2\frac{z_1\cdot (x-y) z_2\cdot (x-y)}{(x-y)^2}\right)^{\ell}}{(x-y)^{2\Delta}}  \\
	&=C_{\Delta,\ell}\frac{\left(z_1\cdot I_{(x-y)}\cdot z_2\right)^{\ell}}{(x-y)^{2\Delta}}\, ,
	\end{aligned}
\end{equation}
Where $I_{(x-y)}^{\mu\nu}$ is the inversion tensor and the constant $C_{\Delta,\ell}$ is a normalisation choice which will be fixed later on. For generic values of $\Delta$, the intertwining kernel is invertible. One can pick choices of normalisation for $C_{\Delta,\ell}$ to try to make these two map the inverse of each other. In this work, we take the simpler convention that this overall constant will be left mostly as is. The reason is to avoid obfuscating when the harmonic decomposition of the kernel becomes singular.

To investigate positivity and irreducibility, we use a version of the partial wave analysis, writing the kernel in a form manifesting the covariance under translation, scaling, and rotations.\footnote{This is an instance of a systematic procedure. The general idea is to perform a harmonic decomposition of the intertwiner over the parabolic subgroup $\widetilde{N}AM$.} From translation invariance, we rewrite the inner product in momentum space
\begin{align}
		\bra{\psi_1}\ket{\psi_2}& =\int \frac{\D^{d}p}{(2\pi)^d}  \ \widehat{\psi_1}^{\dagger}(p)\cdot \widehat{G}_{\overline{\Delta},\ell}(p)\cdot \widehat{\psi_2}(p) \, ,
\end{align}
defining the momentum space intertwiner 
\Bal 
	\widehat{G}_{\Delta,\ell}(p) &= \int \D^{d}x \,  e^{ip\cdot x} \, G_{\Delta,\ell}(x,0)  \\
	&=C_{\Delta,\ell} (z_1\cdot z_2)^{l}\sum_{k=0}^{\ell} \binom{\ell}{k} \left(\frac{2 z_1\cdot \partial_p z_2\cdot \partial_p}{z_1\cdot z_2}\right)^{k} \int \D^{d}x \, \frac{e^{ip\cdot x}}{(x^2)^{\Delta+k}} \, .
\Eal 
The last expression can be evaluated using Schwinger integrals leading to the identity 
\begin{align}
	\int \D^{d}x \,  \frac{e^{ip\cdot x}}{(x^2)^{\Delta}}= \frac{2^{d-2\Delta}\sqrt{\pi}^{d}\Gamma(\frac{d}{2}-\Delta)}{\Gamma(\Delta)}\abs{p}^{2\Delta-d} \, ,
\end{align}
and one can proceeds with the resummations. For convenience, we set the shorthand $h=d/2$, $\nu = \Delta-h$, which are standardly used. Using Leibniz rules twice and reordering the sums, we find  
\Bal
	&(z_1\cdot z_2)^{\ell}\sum_{m=0}^{\ell} \binom{\ell}{m} \frac{\Gamma(\nu-k)}{4^{m}\Gamma(\Delta+m)}
 \left(\frac{2 z_1\cdot \partial_p z_2\cdot \partial_p}{z_1\cdot z_2}\right)^{m} \abs{p}^{-2\nu+2k}\\
 =&\abs{p}^{2\Delta-d} \frac{(-2)^\ell \ \ell!\  \Gamma (\ell-\nu )}{\Gamma (\ell+\Delta)} \left(\frac{z_1\cdot p  z_2 \cdot p}{p\cdot p}\right)^{\ell} P_\ell^{(h-2,\nu-\ell)}\left(\frac{p\cdot p z_1\cdot z_2}{z_1\cdot p z_2\cdot p}-1\right) \, .
\Eal
The function $P_{n}^{a,b}(t)$ is a Jacobi Polynomial. It depends on a single parameter
\begin{align}
	t=\frac{p\cdot p z_1\cdot z_2}{z_1\cdot p z_2\cdot p}-1=\left(z_1\cdot z_2 - \frac{z_1 \cdot p z_2\cdot p}{p\cdot p} \right)\frac{p\cdot p }{z_1 \cdot p z_2\cdot p} \, ,
\end{align}
which measures the angle between $z_1$ and $z_2$ in the plane perpendicular to the vector $p$. To make this effect transparent, we can without loss of generality consider the configuration 
\begin{align}
	p &= \begin{pmatrix}
		\abs{p} \\ 0
	\end{pmatrix}=\abs{p}\widehat{p} ,& z_1 &= \begin{pmatrix}
		+i\sqrt{z^2} \\ z 
	\end{pmatrix}
	,& z_2 &= \begin{pmatrix}
		-i\sqrt{\overline{z}^2} \\ \overline{z}
	\end{pmatrix} ,& z \cdot \overline{z} &=t \sqrt{z^2 \overline{z}^2} \, .
\end{align}
This parametrisation is tuned so that $z_1^{\dagger} \sim z_2$, as happens in the inner product we consider. We have determined 
\begin{align}
	\widehat{G}_{\Delta,\ell}(p)&\equiv \abs{p}^{2\Delta-d} C_{\Delta,\ell} \frac{(-2)^\ell \pi^{h} \ell!\  \Gamma (\ell-\nu )}{\Gamma (\ell+\Delta)}\sqrt{z^2 \overline{z}^2}^{\ell}P_\ell^{(h-2,\nu-\ell)}\left(t\right) \, .
\end{align}

The kernel now makes the invariance under translation ($\widetilde{N}$), dilation ($A$), and rotations ($M$). Taking out the overall factor of $\abs{p}$ which diagonalises the action of dilation, the remaining pieces are defined by the value of the intertwiner for given $\hat{p}^{\mu}$ on $S^{d}$. It must then be decomposable into irreducible representations of the stabiliser group of $p$, which is $SO(d-1)$. This implies a further decomposition of $\widehat{G}_{\Delta,l}(z,\overline{z})$  in a basis of special orthogonal functions associated to the group $SO(d-1)$, known as the zonal harmonics. The statements of positivity and of irreducibility are then statements regarding the harmonic expansion of the intertwining kernel. 

Although the topic of harmonic decomposition is well understood and standard, it is pedagogically worthwhile to review it to setup the logic for when we tackle the spinor case, which is similar but more intricate. We now take a detour to representation theory, before continuing on to the construction of the complementary and exceptional series.

\subsection{Mathematical Interlude : Harmonics Polynomials And \texorpdfstring{$SO(d)$}{SO(d)}}

We have taken a generic function of three vector variables $p$, $z_1$ and $z_2$, and have found it can be recast as a scaling piece and a function of the vectors $z^{\mu}$ and $\overline{z}^\mu$ living in $\mathbb{R}^{d-1}$, which are no more null. The resulting homogeneous polynomial in $z,\overline{z}$ is an encoding of a specific bi-tensor . Such an object can be decomposed on a basis of projectors on definite representations \cite{Bargmann:1977gy,Dolan:2001wg,Dolan:2004up,Dolan:2012wt,Costa:2014rya}.

An irreducible representations of $SO(d-1)$ can be represented by polynomials in the following manner. As any (tensorial) representation lies in the repeated tensor product of the vector representation, we can identify a representation by specifying its projector over a space of tensors of a given number $m$ of indices, $\mathcal{T}_{m}^{(d-1)}$. In this section, we deal with symmetric traceless tensors, whose projectors satisfy
\Bal
	\pi_{m}{}_{\mu_1 \ldots \mu_m}{}^{\nu_1 \ldots \nu_m}&=\pi_{m}{}_{(\mu_1 \ldots \mu_m)}{}^{(\nu_1 \ldots \nu_m)} \, , \\
	h^{\mu_i \mu_j}\pi_{m}{}_{\mu_1 \ldots \mu_m}{}^{\nu_1 \ldots \nu_m}&=0 \, , \\
	\pi_{m \mu_1 \ldots \mu_m}{}^{\sigma_1 \ldots \sigma_m}\pi_{m \sigma_1 \ldots \sigma_m}{}^{\nu_1 \ldots \nu_m}&=\pi_{m}{}_{\mu_1 \ldots \mu_m}{}^{\mu_1 \ldots \mu_{m}} \, .
\Eal
In what follows, we will take all the indices to be perpendicular to $p$. One can simply take $h$ to stand in for the induced metric on the orthogonal plane to $p$, instead of the identity matrix. We will denote the inner product with $h$ using the symbol $a\circ b$, $a^2 = a\circ a$. This projector can be converted into a polynomial, by contracting this projector with generic complex vectors. Using generic vectors $z$ and $\overline{z}$ in $\mathbb{C}^{d-1}$, the resulting object satisfies the properties 
\Bal
	\pi_m(z,\overline{z})&=z^{\mu_1}\ldots z^{\mu_m}\pi_{m}{}_{\mu_1 \ldots \mu_m}{}^{\nu_1 \ldots \nu_m}\overline{z}_{\nu_1}\ldots\overline{z}_{\nu_m} \, , \\
	\pi_m(\lambda z,\rho\overline{z})&=(\lambda \rho)^{m}\pi_m( z,\overline{z}) \, , \\
	\pdv{}{z}\circ\pdv{}{z} \pi_m(z,\overline{z})&=0\, , 
\Eal
Which defines $\pi$ as an homogeneous polynomial, where  tracelessness translates into the projector being a harmonic function.

The upshot is that a given (STT) representation of $SO(d-1)$ is equivalent to a given harmonic homogeneous polynomial. These are called spherical harmonics. The adjective spherical is associated to the homogeneity property, which allows us to fix $\pi$ given its value on the sphere. These conditions uniquely define the spherical harmonics. Start from the ansatz
\begin{align}
	\pi_m(z,\overline{z})=(z^2 \overline{z}^2)^{\frac{m}{2}}f\left(t=\frac{z\circ\overline{z}}{\sqrt{z^2 \overline{z}^2}}\right) \, ,
\end{align}
where $f$ is a polynomial of order $m$ in the variable $t$, harmonicity translates into a differential equation for $f$ 
\begin{align}
	(t^2-1)f''(t)+(d-2)t f'(t)&=l(l+d-3)f(t) ,& f(t)&=c_{(m)}C^{h-\frac{3}{2}}_{m}(t) \, .
\end{align}
Where the factor of $d$ is the dimension of the full-space, matching the rest of this work, not the $p$-orthogonal one, and $C_m$ is the $m$-th Gegenbauer polynomial. We will sometime use $\alpha = h-\frac{3}{2}$ as a shorthand. The normalisation can be fixed by considering the leading order $(z\circ \overline{z})^{m}$ terms which must be set to one, giving 
\begin{align}
	c_{(m)}&=\frac{m!}{2^m\left(h-\frac{3}{2}\right)_{m}} \, , 
\end{align}
Which then fully defines the spherical harmonics.

Spherical harmonics are adapted to decompose a tensor of $SO(d-1)$ of fixed rank into its irreducible components. The problem at hand for us is however slightly different, it is the decomposition of a tensor of $SO(d)$ into its $SO(d-1)$ irreducible pieces. Such a decomposition must encode the branching of representations, and the associated basis are called the zonal harmonics. Indeed, consider a spherical harmonic of $SO(d)$ pulled-back to the space $\mathbb{R}^{d-1}$. The resulting polynomial, although originally traceless, will not be traceless with respect to induced metric. It is then reducible, allowing it to be decomposed into a sum of harmonics of $SO(d-1)$ associated to spins $m\leq l$, multiplied by appropriate trace terms.\footnote{For $d=3$, the projectors over $SO(3)$ are a special form of the Gegenbauer polynomial given by Legendre Polynomials. There, the decomposition into irreducible representations of the subgroup $SO(2)$ is done by Fourier decomposition and one has $-l\leq m \leq l$ instead. For $d=2$, the zonal decomposition is singular since the representations are all one-dimensional.}

To make this explicit, we introduce an external vector $p$ into the discussion, and a bigger inner-product $\cdot$, such that $p\cdot p =1$ and $a\circ b = a\cdot b - a\cdot p \ b\cdot p$. The vector $p$ encodes the breaking of $SO(d)$ into $SO(d-1)$. The breaking of the vector representation gives a breaking of the tensor representation living in its tensor-product, giving rise to projectors $\pi^{\ell}_{m}$ from $\mathcal{T}^{(d)}_{\ell}$ into STT representations with spin-$m$ of $SO(d-1)$, $\ell\geq m \geq 0$, which are called the zonal projectors. By definition, they satisfy 
\Bal
	\pi^{\ell}_{\ell\ }{}_{\mu_1 \ldots \mu_\ell}{}^{\nu_1 \ldots \nu_\ell}&=\pi_{\ell\ }{}_{\mu_1 \ldots \mu_\ell}{}^{\nu_1 \ldots \nu_\ell}  \, , \\
	p^{\mu_1}\ldots p^{\mu_{\ell-m+1}}\pi^{\ell}_{m\ }{}_{\mu_1 \ldots \mu_\ell}{}^{\nu_1 \ldots \nu_\ell}&=0 \, , \\
	\pi^{\ell}_{m\ }{}_{\mu_1 \ldots \mu_\ell}{}^{\sigma_1 \ldots \sigma_\ell}\pi^{\ell}_{m'\ }{}_{\sigma_1 \ldots \sigma_\ell}{}^{\nu_1 \ldots \nu_\ell}&=\delta_{m,m'}\pi^{\ell}_{m\ }{}_{\mu_1 \ldots \mu_\ell}{}^{\mu_1 \ldots \mu_{\ell}} \, ,
\Eal
and acting on a given spin-$\ell$ tensor of $SO(d)$, outputs its spin-$m$ symmetric traceless tensor of $SO(d-1)$ part. 

Using the polynomial encoding, it is straightforward to find an expression for the zonal harmonics. As previously alluded, these projectors must involve the spherical harmonics with some ``trace term" encoding the non-zero contractions in the $p$-direction. We extend the vectors $z$, $\overline{z}$ in the $p$ direction such, allowing us to write objects such as $z\cdot p $, while $z\circ p = 0$ still, as the $\circ$ product is $p$ orthogonal.\footnote{We derive the zonal projectors first as functions of $\mathbb{C}^{d}\times \mathbb{C}^{d}$ initially, which are then restricted to live on the subset of null vectors, $\mathbb{K}^{d}\times \mathbb{K}^{d}$ with $\mathbb{K}^{d}=\{ z \in \mathbb{C}^{d} \mid z\cdot z = 0\}$.} The zonal harmonics are harmonic functions of the reduced laplacian $\partial_z \circ \partial_z$, but not of the full laplacian $\partial_z \cdot \partial_z$ which includes the $p$ direction. Hence, they must be proportional to spherical harmonics for $SO(d-1)$, up to polynomial functions of $z\cdot p$ and $\overline{z}\cdot p$. The following ansatz follows,
\begin{align}
	\pi_{m}^{\ell}(z,\overline{z})&=a_{\ell,m}(z\cdot p \ \overline{z}\cdot p )^{\ell-m}\pi_{m}(z,\overline{z}) , \quad \quad \ell\geq m\geq 0 \, , 
\end{align}
for some normalisation choice $a_{\ell,m}$. The normalisation is fixed by requiring that the harmonics offer a partition of unity on the space of symmetric tensors of $SO(d)$ with $l$ indices $\mathcal{ST}^{(d)}_{\ell}$, 
\begin{align}
	\sum_{m=0}^{\ell}\pi^{\ell}_{m}(z,\overline{z})&= \left(z\cdot z \ \overline{z}\cdot \overline{z}\right)^{\ell/2} \frac{\ell!}{2^\ell(\alpha+\frac{1}{2})_\ell} C_{\ell}^{\alpha+\frac{1}{2}}\left(\frac{z\cdot \overline{z}}{\sqrt{z\cdot z \  \overline{z}\cdot \overline{z}}}\right)  \, .
\end{align}
By specialising to the configuration for which writing $z\cdot z = (z\cdot p)^2$, one can rewrite this as a decomposition of a function of $t$ into a sum of Gegenbauer polynomial, which can be inverted using an integral identity. The output of this computation is the normalisation  
\begin{align}
	a_{\ell,m}=\binom{\ell}{m}\frac{2^{\ell-m}\left(m+\alpha +\frac{1}{2}\right)_{\ell-m}}{\left(2m+2\alpha +1\right)_{\ell-m}} \, .
\end{align}

By construction, the zonal harmonics give an orthogonal decomposition of the set $\mathcal{ST}_{\ell}$. Any map $\tau$ from $\mathcal{ST}_{\ell}$ to itself can then be written as a weighted sum of zonal harmonics 
\begin{align}
	\tau \in \mathcal{ST}_{\ell}^{(d)} \Rightarrow \tau = \sum_{m=0}^{\ell} \kappa_{m}\, \pi^{\ell}_{m} \, .
\end{align}
Such a map is positive definite provided $\kappa_{m}\geq 0$. If some projectors have negative coefficient,  although it is possible to define a positive map by restricting to a subspace orthogonal to the faulty projectors, restricting to this subspace defines a covariant equations of motions \cite{Weinberg:1995mt}. This is not conformally invariant in general. The correct procedure is to write the most general conformally invariant kernel and then consider whether it is positive. One is free to restrict to a subspace only if it is invariant. This can happen if some projectors are absent (simply quotienting by null states), or if there exists a conformally invariant splitting. The latter must be associated to a freedom in the choice of the kernel. 

\subsection{Complementary Series}

We have the tools setup to decompose the intertwiner unto the basis of zonal harmonics. To do so, we consider a specialisation of the previously derived partial wave expansion. If we pick $z\cdot z = 0$, we can identify $z\cdot p = \sqrt{z\circ z \  p\cdot p}$, and similarly for $\overline{z}$. To pull out the factors corresponding to a given harmonic, one can use the completeness relation of the Gegenbauer polynomial. To extract the $m$-th partial wave we need to perform the integral
\begin{align}
	\int_{-1}^{+1}(1-t^2)^{\alpha-\frac{1}{2}}C^{\alpha}_{m}(t) P_\ell^{(h-2,\nu-\ell)}\left(t\right) \, \D t \, ,
\end{align}
which is quite complicated. The tactic we used is to rewrite the Gegenbauer polynomial as a Jacobi Polynomial, and using eq. 7.391.9 of \cite{Gradshtein:2015aa}. The result is an expansion 
\begin{align}\label{eq:boseharm}
	\widehat{G}_{\Delta,\ell}(p) = (p\cdot p)^{\Delta-h}N_{\Delta,\ell}\sum_{m=0}^{\ell} \kappa_{\ell,m}(\Delta)\, \pi^{\ell}_{m}(z_1,z_2,\hat{p}) \, .
\end{align}
Rather miraculously, the final result takes a simple and palatable form. It was first computed in \cite{Dobrev:1977qv}, and is given by  
\begin{align}
	N_{\Delta,\ell} &= \frac{C_{\Delta,\ell} (-1)^l\pi^{h}}{l!}\frac{\Gamma (h-\Delta)}{\Gamma (\Delta-1) (\Delta+\ell-1)} \, ,  & \kappa_{\ell,m}(\Delta)  &=\frac{\left(\overline{\Delta}+m-1\right)_{\ell-m}}{\left(\Delta+m-1\right)_{\ell-m}}\, .
\end{align}

The overall constant $N_{\Delta,\ell}$ must be fixed by picking a normalisation of the kernel. There is no single choice for this normalisation, instead there are various ones suitable for different applications, and often authors have used a panoply of them. As previously mentioned, since the coefficients coming from the Fourier transform might be singular for some values of the scaling dimension, it seems disingenuous to try to tune $C_{\Delta,\ell}$ to remove all of this prefactor, as it encodes genuine information and would hide what happens. Instead we take the much simpler perspective that the only requirement is that the overall factor must be positive (if need be, an overall minus sign can always be adjusted), making us pick a simple normalisation for the 2pt. function 
\begin{align}
	C_{\Delta,\ell}=\frac{1}{(-1)^\ell\pi^h \ell! } \Rightarrow N_{\Delta,\ell} = \frac{\Gamma(h-\Delta)}{\Gamma(\Delta-1)(\Delta+\ell-1)} \, .
\end{align}

At regular points, $N_{\Delta,\ell}$ is not infinite nor zero, and can fix the overall sign provided all the Pochhamer symbols have the same sign. This will define the complementary series representations. The other points are called exceptional, and generate the associated series.

Consider now the situation where $N_{\Delta,\ell}$ is regular.  One must require that all the partial wave coefficients share the same sign, meaning 
\begin{align}
	\frac{\kappa_{m+1}}{\kappa_{m}} = \frac{\Delta+m-1}{\overline{\Delta}+m-1}\geq 0 \, .
\end{align}
The strongest constraint comes from $m=0$, which imposes $\Delta >1$ and $d-1-\Delta>0$. Normalisability is in principle still an issue, which we have to investigate.

For $\ell=0$ there is no such constraint, as there is a single term whose sign can be tuned by rescaling the intertwiner. The only remaining criteria is normalisability of the inner product. The argument is particularly clear in the ``compact'' picture where one can put the origin and infinity on the same footing \cite{Sun:2021thf}. In the non-compact picture we used up to now, one must check that the asymptotic behaviour of the wavefunction is normalisable. Instead, one can work with wavefunctions defined on $\Omega \in S_d$, where the scalar intertwiner takes the form
\begin{align}
	G_{\Delta}(\Omega_1,\Omega_2)=\frac{2^{\Delta}C_{\Delta}}{(1-\Omega_1\cdot \Omega_2)^{\Delta}} \, ,
\end{align}

Which can be obtained by projecting down an embedding space expression \cite{Weinberg:2010ws,Costa:2011wa} or performing an inverse stereographic projection on the position space inner product. Clearly, normalisability of the scalar inner-product for given $\Delta$ implies that of the spinning kernel as well. To know whether this defines a positive function over $S_d \equiv K$, we decompose this bilinear function over the spherical harmonics\footnote{This harmonics decomposition is different from the previous one. Here, we effectively replaced the partial-wave decomposition over the subgroup $\widetilde{N}$ of translation by a decomposition over the larger subgroup $K = SO(d)$.} 
\Bal
	G_{\Delta}&=\frac{C_{\Delta}}{\Gamma(\Delta)}\int_{0}^{\infty} \frac{\D s}{s}\, (2s)^{\Delta}e^{-s+s \Omega_1 \cdot \Omega_2}  \\
	&= C_{\Delta}\Gamma(h-\Delta)\frac{2^{d}\Gamma\left(\frac{d-1}{2}\right)}{4\sqrt{\pi}} \sum_{l} \frac{(d-1+2l)(\Delta)_l}{\Gamma (l+\overline{\Delta})} \,  C_{l}^{\frac{d-1}{2}}(\Omega_1\cdot \Omega_2) \, .
\Eal
Normalisability translates into the positivity of each of these coefficients, which arises provided both $\Delta$ and $\overline{\Delta}$ are positive, i.e. $0\leq \Delta \leq d$. 

This analysis is clearly valid almost everywhere in $\Delta$, and defines the range and properties  of the complementary series $\mathcal{C}_{\Delta,\ell}$. The caveats are the points where the harmonic decomposition is singular, which defines exceptional points, admitting other representations.

\subsection{Understanding Exceptional Points : Scalars} 
\label{sec:exscal}

All representations can be found as subrepresentations of the elementary ones. From our previous analysis, we have found that the representations are irreducible for generic $\Delta$, and become unitary in some restricted range. The caveats of our analysis are the exceptional points, where the representation becomes reducible. Such a distinguished space arises when, for special values of $\Delta$, the intertwiner develops a non-trivial kernel, and the Fourier transform or partial wave expansion becomes singular. This is most easily diagnosed by considering the automorphism generated by performing the shadow transform forward and back, which is caracterised by a constant 
\begin{align}
	G_{\Delta,l}\circ G_{d-\Delta,l} = \mathcal{N}_{\Delta,l} \, ,
\end{align}
which is inversely-proportional to the Plancherel measure of $SO(1,d+1)$ \cite{Knapp:1971aa,Karateev:2018oml,Karateev:2018uk}. It is simply given by  $N_{\Delta,l}N_{d-\Delta,l}$. It is, almost everywhere, a regular function of $\Delta$. It admits poles and zeros at specific integer points, where the isomorphism breaks down into an exact sequence.\footnote{In what follows, we only care about the zeros/irregularities which occur outside of the complementary series. There are spurious poles for $\Delta = h+n$, $n \in \mathbb{N}$, due to the kernel $(x-y)^{-2(h+n)}$ not being conformally invariant \cite{Bzowski:2013sza,Bzowski:2015pba,Bzowski:2018fql}. For points in the complementary series, we can always regulate those by taking limits from the interior of the series, but this is not even needed, nor an issue. At those points, we can use the ultra-local kernel $(-\Box_x)^{n}\delta^{d}(x-y)$ instead. For $n\leq h$, this kernel is still the inverse of $G_{h-n}$, and the isomorphism is preserved. There is no interesting invariant subspace generated by these points. Outside of the complementary range, we must be more careful, as there could be cancellation of zero and poles. For odd $d$, the zero and poles miss each other by half-steps, and we also find there are no interesting invariant subspaces. For even $d$ and $n>h$ however, the points where $G_{h+n}$ becomes ultra-local coincides with the points where $G_{h-n}$ is polynomial and singular, which might cause issue. An explicit computation shows that they still annihilate each-other, and our discussion can still follow through. Effectively, we can perform the analysis in dimensional regularisation keeping $d$ as an external parameter.}
To understand how to deal with this situation and its interpretation, it is useful to consider the case of the scalar operator as an illustrative example of the general structure. For $l=0$, the intertwiner is given by 
\begin{align}
\widehat{G}_{\Delta}(p)&= \frac{\Gamma(h-\Delta)}{\Gamma(\Delta)}\abs{p}^{2\Delta-d} ,&  \mathcal{N}_{\Delta} &= \frac{\Gamma(h-\Delta)\Gamma(\Delta-h)}{\Gamma(d-\Delta)\Gamma(\Delta)} \, .
\end{align}
\begin{figure}[h!]
	\centering
	\includegraphics[width=0.9\linewidth]{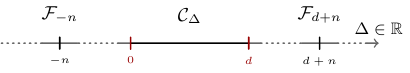}
	\caption{A view of the $\Delta \in \mathbb{R}$ line for scalars. At the centre, there is the unitary complementary series. Outside of its limits in red, we encounter the integer spaced exceptional points.}
\end{figure}

The Plancherel measure has a pole for $\Delta=-n \in -\mathbb{N}$. This is linked to the original Fourier transform. Indeed, the 2pt. function $K_{-n}$ reduces to a polynomial in position space, and in momentum space it becomes a derivative of Dirac delta function
\begin{align}
	G_{-n} \propto (x-y)^{2n} \to \Box_{p}^n \delta^{d}(p)\, ,
\end{align}
the image of which is the set $\Im(G_{-n})\equiv \mathcal{E}_{n}\subset \mathcal{F}_{-n}$ of polynomials of order up to $n$. This forms an invariant subspace of $\mathcal{F}_{-n}$. Since this is a finite dimensional representation and the de Sitter group is not compact, it cannot be unitary except for the trivial case $n=0$ corresponding to the identity. Although very important in some applications with regards to weight-shifting operators \cite{Costa:2011dw,Karateev:2018uk,Costa:2018mcg,Baumann:2019oyu}, we will not consider these further. Likewise, the kernel $\text{Ker}(G_{-n}) \equiv \mathcal{V}_{n} \subset \mathcal{F}_{d+n}$ is also an invariant subspace. These two spaces are candidate representations, which are orthogonal under the natural pairing $(\cdot,\cdot)$, i.e. $\mathcal{E}_n \perp \mathcal{V}_{n}$. 

One might wonder whether their complements give other interesting subspaces. The space $\nicefrac{\mathcal{F}_{d+n}}{\mathcal{V}_{n}}$ is finite dimensional, and made up of functions which are polynomials of degree up to $n$ in $p$,  i.e. this is a finite dimensional space of differential operators. The map from these differential operators to $\mathcal{E}_{n}$ is surjective, linear, and has a trivial kernel ; this makes it a bijective map, making the two spaces isomorphic. This logic follows through for the other quotient space : $\mathcal{V}_n$  can be understood either as the subspace of states in $\mathcal{F}_{d+n}$ for which the naive inner-product is ill-defined, or as the equivalence class of states in $\mathcal{F}_{-n}$ defined up to polynomials. 

This situation can be efficiently encoded in an exact sequence of linear map, which forms the degenerate form of the shadow transform 

\begin{figure}[H]
	\centering
	\includegraphics[width=0.8\linewidth]{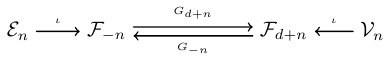}
	\caption{For exceptional points, the shadow isomorphism breaks into a series of exact sequences. Two irreducible spaces survives, and a single one is unitary.}
\end{figure}

Having settled on the study of $\mathcal{V}_n$, this leaves us with a puzzle. Our candidate representation sits in  $\text{Ker}(G_{-n})$, and so we must find a different way to furnish it with an (unitary) inner product. The procedure to do so is systematic, and relies on the quotient space realisation. It answers our two questions

\begin{enumerate}
	\item What is the kernel which defines an invariant inner-product on this irreducible subspace ? This is done through the limiting kernel procedure. It manifest irreducibility, but makes unitarity hard to assess.
	\item Is this kernel unitary ? This is done by relating the inner-product just defined to the one induced by  the (isomorphic) quotient-space picture. 
\end{enumerate} 

We start by finding a suitable kernel. The key is to notice that we have more freedom in choosing the integrand than previously. Under a conformal transformation, the intertwiner is now allowed to transform in-homogeneously; up to a piece proportional to $G_{-n}$. This limiting kernel $\widetilde{G}_{\Delta}$ can be obtained by picking out the (non-trivial) subleading term in the limit to the exceptional point,
\begin{align}
	  \lim_{\Delta \to -n} \left.\frac{G_{\Delta}}{(\Delta+n)}\right\rvert_{\mathcal{V}_{n}} = \lim_{\Delta \to -n} \partial_{\Delta}G_{\Delta}\big\rvert_{\mathcal{V}_n} = \widetilde{G}_{-n} \, .
\end{align}
This operator transforms inhomogeneously under conformal transformation, but it is conformally invariant on the subspace $\mathcal{V}_n$ \cite{Dobrev:1977qv}. We use it to define an inner product
\Bal
	\bra{\psi_1}\ket{\psi_2}&= \left(\psi_1,\widetilde{G}_{-n}\psi_2\right)\, , \, \psi_1,\psi_2 \in \mathcal{V}_{n} \, ,\\
	\widetilde{G}_{-n}(x;y) &\propto (x-y)^{2n}\log(\mu^2 (x-y)^2 ) \, .
\Eal

It is most interesting to note at this point that the natural inner-product is logarithmic. Naively, this seems to signal a breaking of the de Sitter group, as it introduces a scale into the system. This is the source of great a many confusions relating to the discrete/exceptional series. It is then very important to remember that the module we consider is made up of properly smeared states, for which the scale $\mu$ is immaterial. Such a 2pt. function was found while trying to impose a regularisation of the zero mode of a shift-symmetric scalar in de Sitter \cite{Anninos:2023lin}, and is the generally expected result from a  Gupta-Bleuler approach to the quantisation of such a gauge-symmetric scalar \cite{Bros:2010wa,Epstein:2014jaa}. This scale is immaterial and does not break in any way the isometry, as is most clear in the quotient space picture which we will now explain. 

It is naturally quite complicated to directly analyse the unitary properties of this inner product ; but luckily, we do not have to do it like previously. $\mathcal{V}_n$ is isomorphic to the quotient $\nicefrac{\mathcal{F}_{-n}}{\mathcal{E}_{n}}$. Between these spaces, $G_{d+n}$ is bijective, and can be inverted. Since $\mathcal{E}_n \perp \mathcal{V}_n$, any ambiguity in this inverse map leaves the inner-product invariant and can be ignored. Given any choice of inverse map 
\begin{align}
	\varphi : \mathcal{V}_{n} \to \mathcal{F}_{-n} \,  , \, \phi\circ G_{d+n}\big\rvert_{\mathcal{V}_n} = 1 \, , 
\end{align}
it induces an inner-product on $\mathcal{V}_n$ from that of $\mathcal{F}_{-n}$ 
\Bal
	\psi_1,\psi_2 \in \mathcal{V}_n \, , \; \bra{\psi_1}\ket{\psi_2} &= \bra{\varphi(\psi_1)}\ket{\varphi(\psi_2)}_{\mathcal{F}_{-n}}  \\
	&=\big(\psi_1,\varphi\circ G_{d+n}\circ \varphi(\psi_2) \big)= \big( \psi_1, \varphi(\psi_2)\big) \, .
\Eal
The question of unitarity is then transferred onto the quotient space, where we can instead analyse the simpler kernel we have already considered earlier. The limiting kernel procedure produces precisely a construction of an inverse map $\varphi$. This follows directly by taking the limit of the whole shadow automorphism 
\begin{align}
	\lim_{\Delta \to -n}  \frac{G_{d-\Delta}\circ G_{\Delta}}{\Delta+n}\bigg\rvert_{\mathcal{V}_n}= G_{d+n}\circ \widetilde{G}_{-n} \big\rvert_{\mathcal{V}_{n}}=\lim_{\Delta \to -n} \frac{\mathcal{N}_{\Delta}}{(\Delta+n)} = \widetilde{\mathcal{N}}_{-n}\, ,
\end{align}
where the other pieces drop out by virtue of being restricted to the kernel of $G_{-n}$.
 
With this result in hand, we can properly address unitarity, by looking at $\nicefrac{\mathcal{F}_{-n}}{\mathcal{E}_n}$. This is non-trivial, since as we have seen,  although for we can make the (scalar) 2-pt. function seemingly positive for any $\Delta \in \mathbb{R}$ by adjusting the overall coefficient, it is generally not normalisable for $\Delta \geq d$. The representation $\mathcal{V}_n$ will only be unitary if the quotient excises precisely the non-normalisable modes.

To summarise, the space $\mathcal{V}_n \in \mathcal{F}_{d+n}$ is unitary if $G_{d+n}$ is unitary once quotiented by its kernel, and $\mathcal{E}_n$ would be if the analogous statement hold for $G_{-n}$. The inner product would then be given by the limiting kernel, whose unitarity is induced from the quotient space picture. To assess these, it is useful to consider again the compact-picture form of the kernel, which we had expanded over spherical harmonics 
\Bal
	G_{-n}(\Omega_1,\Omega_2) &\propto \sum_{l=0}^{n}(-1)^{l} \frac{(d-1+2l)(n-l+1)_l}{\Gamma(l+d+n)}C^{\frac{d-1}{2}}_{l}(\Omega_1\cdot \Omega_2)\, , \\
	G_{d+n}(\Omega_1,\Omega_2) &\propto \sum_{l=n+1}^{\infty} \frac{(d-1+2l)(n-l+1)_l}{\Gamma(l+d+n)}C^{\frac{d-1}{2}}_{l}(\Omega_1\cdot \Omega_2)\, .
\Eal

We find that $G_{-n}$ only contains harmonics with $l\leq n$ and alternating signs, while the higher ones decouple. The finite-dimensional space of surviving harmonics, $\mathcal{E}_n$, is non-unitary as it should be. The space $\mathcal{V}_n$ is identified with functions on the sphere with harmonics $l>n$. The expression for $G_{d+n}$ shows the converse, that the first $n$ modes decouple, and that the remaining modes appear with positive coefficients, showing it defines a unitary normalisable inner-product. 

Through the relatively intricate construction outlined, we have constructed the scalar exceptional series representations $\mathcal{V}_n$, $n\in \mathbb{N}$. These representations are sometimes called ``exceptional type I'' representations \cite{Sun:2021thf}, to distinguish them from their spinning cousins. We have shown how this module has a logarithmic inner product, which does not induce any issue since the states which are actually part of the representations are those restricted to the physicality conditon ; effectively smearing them. The physicality condition acts like the Gupta-Bleuler quantisation procedure, where one considers a bigger hilbert-space containing unphysical states as well, and one restrict to a physical slice. The seemingly growing modes, associated to polynomial states, are pure gauge, as a consequence of the group-theoretic construction.

\subsection{Spinning Exceptional Series}

Armed with the understanding of the scalar case, we can turn to the more intricate question of generic $\ell$. The spectral function will have poles for $\Delta = d-1+k$, $k\in \mathbb{N}$ again, but the situation is more involved. Indeed, we must differentiate the case $0 \leq t < \ell$, for which $G_{1-t,\ell}$ is not polynomial but still missing projectors, to those with $k = \ell+n$, $0 \leq n$ where, as in the scalar case, $G_{1-\ell-n,\ell}$ is a pure polynomial. This is illustrated in the following figure 

\begin{figure}[H]
	\centering
	\includegraphics[width=1\linewidth]{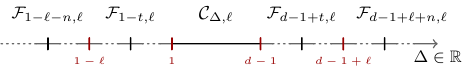}
	\caption{A view of the $\Delta \in \mathbb{R}$ line for tensors. Beyond the complementary series, there are two regions: those lying at integer point point below and above $\ell$.}
\end{figure}

Let us concentrate first on the points at $d-1+\ell+n$, and their cousins at $1-\ell-n$. The scalar discussion is mimicked here, in that we can define the space $\mathcal{E}_{\ell,n} = \Im(G_{1-(\ell+n),\ell})\subset \mathcal{F}_{1-(\ell+n),\ell}$, which is finite dimensional and non-unitary. Likewise, we can define the kernel of this map, $\mathcal{V}_{\ell,n}$, and the two spaces are orthogonal. It is useful to look at the structure of these spaces somewhat more. A generic element of $\mathcal{E}_{\ell,n}$ is a polynomial annihilated by some finite order differential operator, conversely any element of $\mathcal{V}_{\ell,n}$ must have at least a certain number of derivatives associated to it. Exploiting these ideas, these spaces can be written in the form 
\Bal
	\mathcal{E}_{\ell,n} &= \big\{ \text{Polynomials } p(x) \in \mathcal{F}_{1-(\ell+n),\ell} \mid (z\cdot \partial_x)^{n}p = 0  \big\} \, , \\
	\mathcal{V}_{\ell,n} &= \big\{ f(x) \in \mathcal{F}_{d-1+(\ell+n),\ell} \mid \exists g \in \mathcal{F}_{d-1+\ell,\ell+n}\, ,  \; (\mathcal{D}_{z}\cdot \partial_x)^{n}g = f  \big\}  \, .
\Eal

In the first definition, the use of the polarisation $z$ is simply to allow the most general tensor differential equation that annihilates the polynomial. In the second, we use the Todorov operator\footnote{The Todorov operator is an interior derivative to the set $\mathbb{K}^{d}$ of null complex vectors. It is used to free indices which have been contracted with polarisation vectors while ensuring tracelessness of the resulting tensor. See \cite{Bargmann:1977gy,Dobrev:1977qv,Costa:2011wa}.} which releases indices of tensors in the index-free notation, and allowed the elements of $\mathcal{V}_{\ell,n}$ to be the most generic tensorial-function which will be in the kernel. The functional space in which the prefunction $g$ lives can be read up from the scaling and rotational properties of the field and the number of derivatives. In this perspective, we see it is interesting to introduce the two (dual) weight-shifting operators \cite{Dobrev:1977qv}
\Bal
	d &= z\cdot \partial_x , \quad & \quad   \overline{d} &= \mathcal{D}_{z}\cdot \partial_x \, .
\Eal

These differential operators are intertwiners, operators that relates together different representations. In our case, we see they connect the space with integers shifts below and above $\ell$. 

With this in mind, we can consider the doublet of spaces $\mathcal{F}_{1-t,\ell}$ and $\mathcal{F}_{d-1+t,\ell}$, $0 \leq t < \ell$. The map $G_{1-t,\ell}$ is singular as the projectors with $0 \leq m \leq t$ disappear. This allows us to define the invariant subspaces $\mathcal{U}_{\ell,t}=\text{Ker}(G_{1-t,\ell})$ and $\widetilde{\mathcal{V}}_{\ell,t} = \Im(G_{1-t,\ell})$. The space $\mathcal{U}_{\ell,t}$ is analogous to $\mathcal{E}_{\ell,n}$, in that it is annihilated by some finite order differential operator, which ensures it propagates no helicity above $t$, but it is infinite dimensional since its element are not polynomials; it is potentially unitary. 

\begin{figure}[H]
	\centering
	\includegraphics[width=1\linewidth]{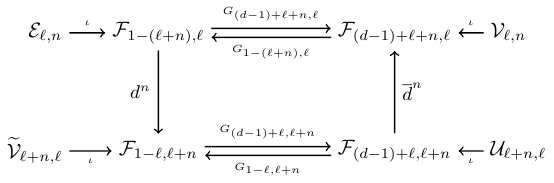}
	\caption{The quartet diagram of \cite{Dobrev:1977qv}, where each directed sequence is exact. It follows there are four potentially interesting irreducible subspaces. A single one is (generically) unitary.}
	\label{fig:quartet}
\end{figure}
\begin{figure}[H]
	\centering
	\includegraphics[width=0.8\linewidth]{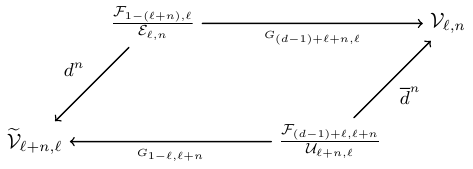}
	\caption{The images of $d^{n}$ and $\overline{d}^{n}$ are $\widetilde{\mathcal{V}}_{\ell+n,\ell}$ and $\mathcal{V}_{\ell,n}$, so they are surjective maps. Taking the quotient, we can make them injective as well. It follows that $\mathcal{V}_{\ell,n}$ and $\widetilde{\mathcal{V}}_{\ell+n,\ell}$ are isomorphic.}
	\label{fig:isomor}
\end{figure}

These new spaces are not totally unrelated to the previous ones however, since weight-shifting operators connects them together. These relations are encoded in a diagram of exact sequences given in fig. \ref{fig:quartet}. For convenience, we have summarised the main rules regarding the interpretation of this type of diagrams in appendix \ref{app:seq}.
\noindent Although it looks quite daunting, these simply express some redundancy in the spaces we defined, showing relations in the different invariant subspaces. Most notably, these imply $\mathcal{V}_{\ell,n}$ is isomorphic to $\widetilde{\mathcal{V}}_{\ell+n,\ell}$, making the latter redundant.\footnote{The reader might wonder what is the goal of this analysis, instead of directly analysing the unitarity of the map $G_{1-t,\ell}$ which induces the inner product on $\widetilde{\mathcal{V}}_{\ell,t}$. The issue is that the harmonic decomposition of $G_{1-t,\ell}$ is quite unwieldy : its first $t+1$ components diverge with positive residue, while the remaining ones are positive. Once we quotient out the modes in $\mathcal{U}_{\ell,t}$, we are restricted to the diverging modes. Circumventing this difficulty is certainly welcome.} This can be seen through the rewriting of the quartet given in fig. \ref{fig:isomor}.

This brings us down to two candidates representations, $\mathcal{V}_{\ell,n}$ which corresponds to the outer points, and $\mathcal{U}_{\ell,t}$ which corresponds to the inner points. For each of these spaces, we can repeat the limiting kernel construction, finding a logarithmic correlator. Unitarity is assessed by recognising that the limiting kernel is an inverse of the map from the quotients. The question then reduces to whether the quotient spaces have positive and normalisable inner products. 

Consider first the space  $\mathcal{V}_{\ell,n}$. Its norm is induced from $G_{d-1+\ell+n,\ell}$, quotiented by its finite dimensional kernel. Through the harmonic decomposition we previously computed, we find 
\begin{align}
	G_{d-1+\ell+n,\ell} \propto \sum_{m =0}^{\ell} (-1)^{\ell-m}\frac{(n+1)_{\ell-m}}{(d-1+\ell+n)_{\ell-m}}\pi^{\ell}_{m} \, .
\end{align}
All the modes up to $\ell$ survive, but with alternating coefficients.\footnote{This might seem counterintuitive, since we previously claimed that this map should have a non-trivial kernel. The point is that the kernel is located not in the intrinsic spin, but in the orbital spin. To make it manifest, one would have to repeat our discussion but in the compact-picture, doing the harmonic analysis over $S_d$, as we did for the scalar. This is however not particularly enlightening, since the explicit position space form of the reciprocal maps makes the presence of the kernel completely transparent already.} This map is not positive, and so the representation $\mathcal{V}_{\ell,n}$ is not unitary, except for $\ell=0$ where it reduces to $\mathcal{V}_n$.

For the space $\mathcal{U}_{\ell,t}$, we have to look at $G_{d-1+t,\ell}$ instead. Only the partial waves with helicities $t < m \leq l$ survive, and they all appear with positive coefficient
\begin{align}
	G_{d-1+t,\ell} &\propto \sum_{m=t+1}^{\ell} \frac{(\ell-1-t)!}{(m-1-t)!(d-1+\ell+t)_{\ell-m}}\pi^{\ell}_{m} \, .
\end{align}
The inner product is then positive, but it remains to check whether it is finite over our representation. Proofs can be found in \cite{Dobrev:1977qv,Sun:2021thf}. As a quick argument, one can note that  the asymptotic behaviour of the spinning wave functions is not worse than the scalar ones, which were normalisable at these exceptional points as seen from the compact picture. 

The representation $\mathcal{U}_{\ell,t}$, $t=0, 1,\ldots \ell-1$ is a depth-$r$, $r=(\ell-t)$ partially-conserved tensor representation, since it contains a tensor field which is annihilated by taking $r$ divergences. It is sometimes called ``Exceptional type II'' representation, or spinning exceptional series.

\section{Story of Fermi : Rarita-Schwinger And  Higher}\label{sec:spinor}

We have seen how to construct bosonic representations by considering the positivity of a partial wave decomposition of the intertwining kernel. In this section, we generalise this to fields which transform as tensor-spinors. This requires us to redevelop some less common machinery, related to Clifford Analysis.

Once this setup is done, we are in a position to make our first result, showing that there are no complementary or exceptional series representation for tensor-spinors. Beautifully, this boils down to a simple fact, that $(\gamma\cdot p)^2 = p\cdot p >0$, and so eigenvalue comes in pairs. Since the harmonic decomposition used is only valid for $d>3$, lower-dimensional cases require a more specific formalism. We relegate the case $d=2,3$ to the next section. There, we find that there exist unitary exceptional spinor representations for $d=3$ only.

\subsection{Statement of The Problem}

The goal is now to extend our range of representations by picking states to transform under the subgroup $M$ under representation $\rho$ of the $Spin(d)$ group. The most straightforward generalisation of the symmetric traceless tensor is the symmetric-tensor spinor with $\ell$ indices, the bulk counterpart of which are generalised Rarita-Schwinger fields. We will call such objects higher-spinors or spinor-tensors. 

In more details, we considers the wavefunctions $\psi_{\Delta}^{\mu_1 \ldots \mu_\ell}(x)$ transforming as primaries with scaling dimension $\Delta$ which are spinor-valued. One can again define an index-free formalism. For irreducibility, we impose
\begin{align}
	&\begin{aligned}
		\delta_{\mu_i\mu_j}\psi_\Delta^{\mu_1 \ldots \mu_\ell}&=0 \\ 
		 \gamma_{\mu_i}\psi_\Delta^{\mu_1 \ldots \mu_\ell}&=0
	\end{aligned} & &\begin{aligned}
		\psi_\Delta^{(\mu_1 \ldots \mu_\ell)}&=\psi_\Delta^{\mu_1 \ldots \mu_\ell}\\
		\psi_{\Delta}(x,z)&=z_{\mu_1} \ldots z_{\mu_\ell} \psi_{\Delta}^{\mu_1\ldots \mu_\ell}
	\end{aligned}
\end{align}
The task of constructing an inner product reduces again to that of finding a conformally covariant 2pt. functions for fields of this type, 
\begin{align}
	\bra{\psi_1}\ket{\psi_2} = \int \D^{d}x \, \D^{d}y \, \overline{\psi_1}(x)\cdot \mathcal{G}_{\overline{\Delta},\ell}(x,y)\cdot \psi_2(y) \, ,
\end{align}
where $\mathcal{G}_{\overline{\Delta}}$ is now Clifford valued. Since we are working with euclidean spinors, the Dirac conjugate is simply given by taking the hermetian conjugate.

As is always the case, there exists a principal series of UIR with $\Delta = \frac{d}{2}+i\nu$, as the $L^2$ inner product is unitary and irreducible.\footnote{Strictly speaking, the intersection with the real axis is reducible \cite{Hirai:1962aa,Dobrev:1977qv}, which does not happen for bosons. This is simply due to the quotient of the null descendant induced by the Dirac equation, as we comment on later.} The bulk interpretation of the scaling dimension is slightly different for fermions, as the quantisation of massive fermions yields $\Delta = \frac{d}{2}+im$, with $m$ the renormalised, Lagrangian mass of the field, which must be real for a real lagrangian \cite{Pethybridge:2021rwf,Schaub:2023scu,Letsios:2023qzq}. This in itself is not enough to discard more exotic representations, since they could ultimately be realised in non-Lagrangian scenarios.

To investigate the potential complementary or exceptional series, we need to pick $\Delta \in \mathbb{R}$ and look at the non-trivial 2pt. function of spinor-tensors. These can be built using standard CFT techniques \cite{Iliesiu:2015qra,Iliesiu:2017nrv,Isono:2017grm}, 
\Bal
	\mathcal{G}_{\Delta,\ell}(x,y)=C_{\Delta,\ell+\frac{1}{2}}\frac{\left(z_1\cdot I_{(x-y)} \cdot  z_2 \right)^{\ell-1}}{(x-y)^{2\Delta+1}}\big(&\gamma\cdot (x-y)\left(z_1\cdot I_{(x-y)} \cdot  z_2 \right) \\
	 &+\frac{\ell}{d-3+2\ell}\gamma\cdot z_1 \gamma\cdot (x-y)\gamma\cdot z_2\big) \, .
\Eal
As before $I_{(x-y)}^{\mu\nu}$ is the inversion tensor appearing in the 2pt. function of vector fields. The first term is mandated from conformal symmetry, while the second one is simply a subtraction to ensure $\gamma$-tracelessness.

This kernel is already much more intricate than its bosonic counterpart. Before delving into the Fourier transform and subsequent partial wave decomposition, let us gain some intuition for what the fermionic nature of the field implies. Consider  $\ell =0$, i.e. the Dirac spinor. In momentum space, the kernel reduces to
\Bal
	\widehat{\mathcal{G}}_{\Delta,0}(p)&= p^{2\Delta-d} C_{\Delta,\frac{1}{2}}\frac{4^{h-\Delta}\pi^{h}\Gamma(h-\Delta-\frac{1}{2})}{2\Gamma(\Delta+\frac{1}{2})}\gamma\cdot \hat{p} \\ 
	&=  p^{2\Delta-d} C_{\Delta,\frac{1}{2}}\frac{4^{h-\Delta}\pi^{h}\Gamma(h-\Delta-\frac{1}{2})}{2\Gamma(\Delta+\frac{1}{2})}\left(P_{+}-P_{-} \right) \, .
\Eal
This expression does not form a positive definite inner product. The issue with complementary series spinor is glaring : the matrix $\gamma \cdot\hat{p}$ possesses two eigenvalues, $+1$ and $-1$, hence once decomposed over projectors $\gamma\cdot \hat{p} \, P_{\eta} = \eta P_{\eta} $, it always gives rise to an indefinite  inner product on the representation. 

The only way out is to kill the projection, by imposing $\gamma\cdot p\  \psi = 0 $. This translates into a covariant equation of motion for the wavefunctions \cite{Weinberg:1995mt}, restricting us to a short-multiplet, which enforces it to be a free massless Dirac field. This field lives at the bottom of the principal series, for $\nu = m =0$. Clearly, this fields possesses null descendant, which must simply be quotiented out quite trivially.

In what follows, we will show that this is precisely what happens for higher-spinors as well. For $d>3$, we will find the only possible way out is to restrict to wave functions satisfying $\gamma\cdot p \ \psi = 0$.  However these are precisely the free fields satisfying the Dirac equation and for which $\Delta = \frac{d}{2}$, that sits on the tail of the principal series. 

To make this statement concrete, we must proceed with the construction of the harmonic decomposition. As we did previously, we need to compute the momentum space kernel. It follows from the bosonic kernel by taking one further derivative and shifting the weight. Luckily, one does not actually need the explicit expression, simply the schematic expression, 
\begin{align}\label{eq:schematic}
	\widehat{\mathcal{G}}_{\Delta,\ell} \approx (p\cdot p)^{\Delta-h}\left( \gamma\cdot \hat{p}\ldots + \gamma\cdot \hat{p} \, \gamma_{\mu\nu}z^{\mu}\overline{z}^\nu \ldots + \gamma\cdot z \ldots + \gamma \cdot \overline{z} \ldots   \right)
\end{align}
With $z$ and $\overline{z}$ the polarisations projected in the orthogonal plane to $\hat{p}$. Each term can be written in term of some differential operator acting on Jacobi polynomials, but their precise form is in the end, not needed.

The crucial matter is clearly, how to decompose an object such as eq. \eqref{eq:schematic} into projectors; into irreducible pieces with respect to the stabiliser group $Spin(d-1)$. The conceptual issue is that one must define the spinorial equivalent of the harmonics, which is not as well known as the bosonic analog. This is the goal of the next section.

\subsection{Intermezzo : Clifford Harmonics And \texorpdfstring{$Spin(d)$}{Spin(d)}}

We have seen how the harmonic decomposition story goes for bosonic representations in much details. The logic has been setup to do the same for fermions. We identify representations by their projectors. The technicality arises in that there is a family of four type of projectors to consider to have a full  partition of unity on the space of spinor-tensors. 

We are interested in performing a harmonic decomposition of an object which transforms as a symmetric tensor taking value not in the field of real or complex numbers as previously, but in the Spinor representation. We will consider vector indices in $SO(d-1)$, and the associated projector $\Pi$ must take value in the Clifford algebra $\mathcal{C}\ell_{d-1}$. We first look at the irreducible pieces. The central issue one has to face is that such irreducible representations are of two types, the $\left[m+\frac{1}{2}\right]$ irreps, which we call traceless, and $\left[m-\frac{1}{2}\right]$ pieces, which is conceptually similar to a zonal component already, and that we call traceful. The first one is $\gamma$-traceless, while the second one is a pure-gamma trace, and is fixed by requiring that the sum of the two give a partition of the tensors of spin $m$. All of this is simply the projector reformulation of the well known factors from the study of the tensor product of spinor and vector representations of $SO(d-1)$. We thus define the projector
\Bal
	\Pi^{m}_{\pm}{}_{\mu_1 \ldots \mu_m}{}^{\nu_1 \ldots \nu_m}&=\Pi^{m}_{\pm}{}_{(\mu_1 \ldots \mu_m)}{}^{(\nu_1 \ldots \nu_m)} \, , \\
	\delta^{\mu_i \mu_j}\Pi^{m}_{\pm}{}_{\mu_1 \ldots \mu_m}{}^{\nu_1 \ldots \nu_m}&=0 \, .
\Eal
Consider then the generic $\left[m+\frac{1}{2}\right]$ ``traceless'' case, with $\Pi^{m}_{+}$, which satisfies 
\begin{align}
	\gamma^{\mu_i}\Pi^{m}_{+}{}_{\mu_1 \ldots \mu_m}{}^{\nu_1 \ldots \nu_m}&=0 \, .
\end{align}
Meanwhile, the $-$ ``traceful'' projector will not be annihilated by the contraction, and it will come together with $\Pi^{m}_{+}$ to completely span the space of symmetric-traceless-spinor. This translate into the requirement that the traceless and traceful parts sum into the tensorial projectors,\footnote{This type of equation is known in the Clifford analysis literature as a Fischer decomposition \cite{Clif:Van-Lancker:2001aa,Clif1}. It encodes that harmonics for spinors form a refinement of spherical harmonics.}
\begin{align}
	\pi^{m}{}_{\mu_1\ldots \mu_m}{}^{\nu_1 \ldots \nu_m}&=\Pi^{m}_{+}{}_{\mu_1 \ldots \mu_m}{}^{\nu_1 \ldots \nu_m}+\Pi^{m}_{-}{}_{\mu_1 \ldots \mu_m}{}^{\nu_1 \ldots \nu_m} \, .
\end{align}

As we did in the bosonic case, these projectors can be encoded as a (matrix valued)-polynomial by contracting its indices with polarisations. The projectors satisfy the properties 
\Bal
	\Pi^m_{\pm}(z,\overline{z})&=z^{\mu_1}\ldots z^{\mu_m}\Pi^{m}_{\pm}{}_{\mu_1 \ldots \mu_m}{}^{\nu_1 \ldots \nu_m}\overline{z}_{\nu_1}\ldots\overline{z}_{\nu_m} \, , \\
	\Pi^{m}_{\pm}(\lambda z,\rho\overline{z})&=(\lambda \rho)^{m}\Pi^m_{\pm}(z,\overline{z})\, , \\
	\pdv{}{z}\circ\pdv{}{z} \Pi^m_{\pm}(z,\overline{z})&=0 \, , \\
	\gamma\circ \pdv{}{z}\Pi^{m}_{+}(z,\overline{z})&=0 \, .
\Eal
The first three conditions are unchanged from before. The fourth condition is a stronger requirement than the third one, and implies it.\footnote{In the Clifford analysis literature, see for example \cite{Clif1,Clif2,Clif:Delanghe:1992aa,Clif:Van-Lancker:2009aa,Clif:Van-Lancker:2015aa}, this null-Dirac condition is called (left) monogenicity. Clifford analysis is concerned with the study of monogenic functions. The term monogenic stems from the equation being a single derivative constraint. The theory of monogenics generalises the study of harmonics, while offering parallels to the theory of holomorphic functions.} We will refer to such function as Clifford harmonics. They tell us that a given representation of the spinor-tensor, specified through its projector, is equivalent to a homogeneous Clifford polynomial. Such objects we call Clifford spherical harmonics. This construction forms the $Spin(d)$ analogue of our construction of the $SO(d)$ representations.

From basic properties of Clifford algebra, we know the projector $\Pi^m$ can be expanded on a basis,
\begin{align}
	\Pi_m(z,\overline{z})&= \Pi^{0}_{m}+\gamma^{\mu}\Pi^{1}_{m,\mu}+\gamma^{\mu\nu}\Pi^{2}_{m,\mu\nu} \, ,
\end{align}
where each of the $\Pi^{k}_{m}$ are referred to as subprojectors, and are valued in the identity element of the Clifford algebra. The subprojectors must be totally antisymmetric homogeneous polynomial of degree $m$ in the variables $z$ and $\overline{z}$. This implies directly that the expansion in  subprojectors truncate at second order in the Clifford algebra. 

\paragraph{Remark on the dimension dependence:}{We will pick $\gamma_\mu \in \mathcal{C}\ell_{d}$ but still take  $\mu$ to live in $d-1$ dimensions, since we ultimately want to work with zonal harmonics. As is often the case regarding spinors, care must be given to the number of dimension one works in. If the dimension is odd or even, the chiral matrix can be incorporated as follows
\begin{itemize}
	\item If $d=2k$, $\gamma_\star$ could be accommodated by multiplying by $\frac{1\pm \gamma_\star}{2}$.
	\item If $d=2k+1$, we can identify $\gamma\cdot p$ with the $\gamma_\star$ of $\mathcal{C}\ell_{d-1}$.
\end{itemize}
For low enough dimension, the algebra degenerates and one cannot separate the zonal decomposition and the decomposition onto definite $\gamma\cdot p$ eigenvectors
\begin{itemize}
	\item For $d=3$, the Pauli matrices satisfy $\gamma^{\mu\nu}=\epsilon^{\mu\nu}\gamma_3$, and we can without loss of generality pick $\gamma\cdot p = \gamma_3$. 
	\item For $d=2$, there is the exceptional simplification  whereby $\gamma^{\mu\nu}=0$, and the algebra fully degenerates to level one.
\end{itemize}
}
Back to the projectors. Using the scaling and Lorentz properties of $\Pi$, we can write ansätze for the subprojectors, 
\Bal
	\Pi^{0}_m &= (z^2\overline{z}^2)^{\frac{m}{2}}f_0(t) \, ,\\
	\Pi^{1}_{m,\mu}&=(z^2\overline{z}^2)^{\frac{m}{2}} \left( \frac{z_\mu}{\sqrt{z^2}}f(t)+ \frac{\overline{z}_\mu}{\sqrt{\overline{z}^2}}g(t)\right)  \, , \\
	\Pi^{2}_{m,\mu\nu}&=(z_\mu\overline{z}_\nu-z_\nu\overline{z}_\mu)(z^2\overline{z}^2)^{\frac{m-1}{2}}f_2(t) \, .
\Eal
\ridoff{
Before going to the next step, it is useful to note a few formula related to the $\gamma$-matrices : 
\Bal
	\gamma_\mu\gamma_\nu &= \gamma_{\mu\nu}+\eta_{\mu\nu} \, , & \gamma_{\mu}\gamma_{\rho\lambda}&=\gamma_{\mu\rho\lambda}+\delta_{\mu\rho}\gamma_\lambda-\delta_{\mu\lambda}\gamma_\rho\, , \\
	\gamma\cdot \gamma &= d-1 \, , & \gamma_\mu \gamma^{\mu\nu}&=(d-2)\gamma^{\nu}\, , \\
	\gamma_{\rho\lambda}\gamma_{\mu}&=\gamma_{\mu\rho\lambda}-\delta_{\mu\rho}\gamma_\lambda+\delta_{\mu\lambda}\gamma_\rho \, , & \gamma^{\mu}\gamma_{\mu\rho\lambda}&=(d+1)\gamma_{\rho\lambda}\, ,
\Eal
and the more involved
\begin{align}
	\gamma_{\mu\nu}\gamma_{\sigma\lambda}=\gamma_{\mu\nu\sigma\lambda}-\delta_{\mu\sigma}\delta_{\nu\lambda}+\delta_{\mu\lambda}\delta_{\nu\sigma}-\delta_{\mu\sigma}\gamma_{\nu\lambda}+\delta_{\mu\lambda}\gamma_{\nu\sigma}+\delta_{\nu\sigma}\gamma_{\mu\lambda}-\delta_{\nu\lambda}\gamma_{\mu\sigma} \, .
\end{align}
}

Let us investigate the traceless projector. This gives the differential equation 
\begin{align}
	\slashed{\partial}_z \Pi_m =0 &=  \partial^\mu \Pi^{1}_\mu+\gamma^{\mu}\left(\partial_\mu \Pi^{0}+2\partial^\nu\Pi_{\nu\mu}^{2}\right)+\gamma^{\mu\nu}\partial_\mu \Pi^{1}_\nu \, ,
\end{align}
which translates into a set of coupled differential equations for the functions $f_0, f_2, f$ and $g$ 
\Bal
	0&=(d+m-2) f-\left(t^2-1\right) g'+m t g  \, , & 0&=f'+t g'-m g \, , \\
	0&=2\left(t^2-1\right)f_2'+2(m-1) tf_2-tf_0'+m f_0 \, , & 0&=2(d+m-3) f_2+f_0' \, .
\Eal
The last equation can be inputted back into the third one to give a second order differential system for $f_0$, which turns out to be the $m$-th Gegenbauer Polynomial defining equation again, 
\begin{align}
	f_0(t)&= \frac{2\alpha+m}{2(\alpha+m)}c_{(m)}C^{\alpha}_{m}(t) \, , & f_2(t)&= -\frac{\alpha c_{(m)}}{4\alpha+m}C^{\alpha+1}_{m-1}(t) \, .
\end{align}
This however, does not fix the $f$ and $g$ functions, and this is actually a feature of the problem at hand. We postpone this issue slightly; let us require that the projector be traceless from the right as well, which gives the same set of equation but with the role of $f$ and $g$ interchanged. One can check that the resulting equations are not compatible with each others, setting $\Pi^{1}=0$. 

Taking a step back, these results make sense when considering the normalisation of the projectors. If $z^2=0=\overline{z}^2$, we must reduce to an expression which looks like $(z\circ \overline{z})^m$ minus the $\gamma$-traces, which can be computed using the Todorov operator. Going through this exercise, we find the normalisation condition
\Bal
	&(z\circ \overline{z})^{m-1}\left(z\circ \overline{z} - \frac{m}{d-3+2m} \gamma\circ z \gamma\circ \overline{z} \right)\\
	=&\frac{d-3+m}{d-3+2m}(z\circ \overline{z})^{m-1}\left( z\circ \overline{z}-m\gamma_{\mu\nu}z^{\mu}\overline{z}^{\nu} \right) \, .
\Eal
Clearly, the projector for the $m+\frac{1}{2}$ spin component, will have no solitary $\gamma_\mu$ term. The previous solution for $f_0$ satisfies precisely this normalisation. 

Now, armed with insights, we can look back at the previous construction and understand the general structure faster. Indeed
\Bal
	\Pi^{m}_{++}&=\left(1-\frac{1}{2(\alpha+m)}z\circ \gamma\gamma\circ \partial\right)\pi_{m} \\
	&=\pi_{m}\left(1-\frac{1}{2(\alpha+m)}\overset{\leftarrow}{\overline{\partial}}\circ \gamma \gamma\circ \overline{z} \right) \, ,
\Eal
is $\gamma$-traceless on both sides by construction, and satisfies all the properties of a Clifford harmonic. Moreover, this expression makes it easy to show that 
\Bal
	\Pi^{m}_{++}(z,\partial)\Pi^{m}_{++}(z,\overline{z})&=\left(1-\frac{1}{2(\alpha+m)}z\circ \gamma\gamma\circ \partial\right)^2\pi_m(z,\partial)\pi_{m}(z,\overline{z}) \\
	&= \Pi^{m}_{++}(z,\overline{z}) \, ,
\Eal
making it a good projector. An explicit computation shows it to be identical to the one given previously. In this form however, the generalisation to find the traceful projector, and the zonal Clifford harmonics is trivial,
\Bal
	\Pi^{\ell,m}_{++}&=\left(1-\frac{1}{2(\alpha+m)}z\circ \gamma\gamma\circ \partial\right)\pi^{\ell}_{m} \, ,\\ 
	\Pi^{\ell,m-1}_{--}&=\left(\frac{z\circ \gamma\gamma\circ \partial}{2(\alpha+m)}\right)\pi^{\ell}_{m} \, .
\Eal
These projectors enjoys the following expected properties, that follow from those of the gamma matrices and zonal harmonics 
\Bal
	\Pi^{\ell,m}_{\alpha \alpha}(z,\partial)\Pi^{\ell,m'}_{\beta\beta}(z,\overline{z})&=\delta_{m,m'}\delta_{\alpha \beta}\Pi^{\ell,m}_{\alpha \alpha}(z,\overline{z}) \, , \\
	\sum_{m=0}^{\ell} \Pi^{\ell,m}_{++}(z,\overline{z})+\Pi^{\ell,m}_{--}(z,\overline{z})&= \left(z\cdot z \ \overline{z}\cdot \overline{z}\right)^{\ell/2} \frac{\ell!}{2^\ell(\alpha+\frac{1}{2})_\ell} C_{\ell}^{\alpha+\frac{1}{2}}\left(\frac{z\cdot \overline{z}}{\sqrt{z\cdot z \  \overline{z}\cdot \overline{z}}}\right) \, .
\Eal
For later convenience, we defined the traceful projector with a shift in the indices, so that $\Pi^{\ell,\ell}_{--}=0$, and $\Pi^{\ell,-1}_{--}=0$ of course. In this convention, it always picks the component with spin $\left[m+1/2\right]$, and there is a single component with spin $\left[\ell+\frac{1}{2}\right]$. The definitions are chosen so that both $\Pi^{\ell,m}_{\pm\pm}$ projects out a component which has spin $\left[m+ \frac{1}{2}\right]$ out of the reduction of $ [\ell] \otimes \left[\frac{1}{2}\right]$.

There is yet another expression for $\Pi_{++}^{l,m}$ which is useful in derivations. Since the Dirac operator squares to the Laplacian, we can sandwich the zonal harmonics with appropriate factors to retrieve the projector just given. Through an explicit computation one finds 
\begin{align}
	\Pi^{l,m}_{++}&= \frac{(2 \alpha +l+m+1) (2 \alpha +l+m+2)\gamma^\mu\gamma^{\nu}}{4 (l+1) (2 \alpha +2 l+1) (\alpha +m+1)^2}\pdv{}{z^{\mu}}\pdv{}{\overline{z}^\nu}\pi_{l+1,m+1}(z,\overline{z}) \, 
\end{align}

The projectors just obtained are certainly an allowed solution to the defining equations, but it does not fully solve our problem. In principle, the $f$ and $g$ functions can be non-zero, for a projector which is traceless on one side, traceful in the other. This arises from the group-theoretical fact that in the harmonic decomposition of $ \left[\frac{1}{2}\otimes j\right]^{spin(d-1)} $ there are multiple factors of the same representation $\left[m+\frac{1}{2}\right]$ and $\left[(m+1)-\frac{1}{2}\right]$ which arise from the $++$ or $--$ sector. Since there are two projectors picking out the same group representations, there is an ambiguity in their action. This implies the existence of a map that transforms one into the other. This is provided by the two interpolating projectors $\Pi^{l,m}_{+-}$ and $\Pi^{l,m}_{-+}$, which involve the non-trivial $f$ and $g$ factors, and a single $\gamma$-matrix.

Directly solving the equations for $f$ and $g$ is hard to interpret, as the interpolating projectors are only relevant in the zonal expansion. It is more efficient to consider building them as differential operators acting on the zonal spherical harmonics. A satisfying solution is found to be given by 

\Bal
	\Pi^{l,m}_{+-}&= \sqrt{\frac{2 \alpha +l+m+1}{l-m}} \frac{z\cdot p}{2(\alpha+m+1)} \ \gamma \circ \partial\pi_{l,m+1} \, , \\
	\Pi^{l,m}_{-+}&=\pi_{l,m+1}\gamma \circ \overset{\leftarrow}{\overline{\partial}}\ \frac{\overline{z}\cdot p}{2(\alpha+m+1)} \sqrt{\frac{2 \alpha +l+m+1}{l-m}} \, .
\Eal
Note that by our previous definitions, $\Pi_{\pm,\mp}^{l,l}=0$, as there is only a single way to pick out a spin $\left[l+\frac{1}{2}\right]$ component out of the reduction of $\left[l \right]\otimes \left[\frac{1}{2}\right]$. One can derive straightforwardly the full set of identities encoding the algebra of these projectors 
\begin{align}
	\Pi^{l,m}_{\alpha \beta}(z,\partial)\Pi^{l,m'}_{\sigma\lambda}(z,\overline{z})&=\delta_{m,m'}\delta_{\beta \sigma}\Pi^{l,m}_{\alpha \lambda}(z,\overline{z}) \, ,
\end{align}
and we note that these operators can further be dressed by multiplying by appropriate factors of the projector $P_{\pm}=\frac{1\pm \cdot p}{2}$ over definite eigenvalue of $\gamma\cdot p$.

\subsection{No Complementary Nor Exceptional Fermions For $d>3$}

From the form of the different projectors and given the schematic form of the kernel in eq. \eqref{eq:schematic}, one can see how the harmonic decomposition proceeds,
\Bal
	\overline{\psi}\cdot\widehat{\mathcal{G}}_{\Delta,\ell}\cdot\psi &\propto\sum_{m=0}^{\ell} \overline{\psi}\left(\gamma\cdot{p}\left(a_{\ell,m}\Pi_{++}^{\ell,m}+d_{\ell,m}\Pi_{--}^{\ell,m}\right)+b_{\ell,m}\Pi_{+-}^{\ell,m}+c_{\ell,m}\Pi_{-+}^{\ell,m} \right)\psi \\
	&= \sum_{m=0}^{\ell}\sum_{\eta = \pm} \begin{pmatrix}
		\overline{\psi}\Pi^{\ell,m}_{+-}P_{\eta} & \overline{\psi}\Pi^{\ell,m}_{--}P_{\eta}
	\end{pmatrix}\begin{pmatrix}
		-a_{\ell,m} \eta & b_{\ell,m} \\ c_{\ell,m} & d_{\ell,m} \eta 
	\end{pmatrix}\begin{pmatrix}
		P_{\eta}\Pi^{\ell,m}_{-+}\psi \\
		P_{\eta}\Pi^{\ell,m}_{--}\psi
	\end{pmatrix} \\
	&=\sum_{m=0}^{\ell}\sum_{\eta = \pm}  \overline{\Psi}_{\ell,m,\eta}M_{\ell,m,\eta}\Psi_{\ell,m,\eta} \, .
\Eal
This is the analog of the bosonic harmonic decomposition in eq. \eqref{eq:boseharm}. Positive-definiteness is now a matrix constraint, which requires that all the $M_{\ell,m,\eta}$ matrices have eigenvalues all of the same sign. This is a stringent requirement, which is much harder to satisfy than the equivalent bosonic statement. In fact, this is really an impossible requirement, independently of the exact form of each of the coefficients $a_{\ell,m},b_{\ell,m},c_{\ell,m},d_{\ell,m}$.

Consider the $m=\ell$ contribution, 
\begin{align}
	\sum_{\eta}\overline{\Psi}_{\ell,\ell,\eta}M_{\ell,\ell,\eta}\Psi_{\ell,\ell,\eta}&= \sum_{\eta=\pm} \eta \left(a_{\ell,\ell}\overline{\psi}\Pi^{\ell}_{++}P_{\eta} \psi\right) \, .
\end{align}
These two contributions from the leading $l+1/2$ representation degenerate into single block, which always comes in a pair with opposite signs. This means the inner product is always indefinite. There is no constraint with can be put on the field which would make the inner product sign definite, hence there is no salvaging the complementary series. This argument repeats identically for all helicities. Since each $2\times 2$ matrices appears in pairs with identical determinant but opposite trace, it must be that the eigenvalue flip sign with $\eta$. This traces back simply to the sign indefiniteness of $\gamma\cdot p$, as in the scalar case. 

We have shown in general that there are no complementary series spinors. Could there be exceptional ones ? If one could separate out the modes with opposite norms, yes. This is not possible in general dimension however, precisely because of the form of the decomposition : since the eigenvalues of $M_{\ell,m,\eta}$ for given $\ell,m$ and are independent of $\eta$ apart from an overall sign, singular-points cannot select out modes with definite signs. To be able to single them out would imply an invariant equation of the form $\gamma\cdot \partial \psi = (\ldots) \psi$, which is not conformally invariant except for the vanishing case. This reduces back to the ``massless'' spinors, with $\Delta = \frac{d}{2}$.

An alternative understanding of this impossible splitting of modes with opposite helicities can be gathered by thinking of special conformal transformations. Consider the Dirac spinor, under special conformal transformation it behaves like a two-level system (the positive and negative helicities), and the rotational part $\sim \gamma_{\mu\nu}$ mixes them together. This features persist for higher-spinor, where the opposite modes, which are simply related to the two helicities of the Dirac spinor tensor-product with the tensor modes, mix together under special conformal transformation.

In conclusion, we have shown that there are no unitary complementary or exceptional spinor-tensor representations for $d>3$. The caveats of our discussion is that our harmonic decomposition degenerates for low-dimensions. Indeed, the Clifford algebra contracts down as we reduce the dimension, and the structure $\gamma_{\mu\nu}$ in the harmonic projectors start to interact with $\gamma\cdot p$, mixing together $l,m$ and $\eta$. Hence, it is only for $d=1,2,3$ that one can hope to find exceptional spinorial representations, beyond the one encoded by fields satisfying the Dirac equation. These lower-dimensional discussions are the topic of the next section.

\section{Peculiarities of Lower Dimensions}\label{sec:dim}

We have found from our general analysis that in $d>3$ spinors must reside in the principal series. The goal of this section is to perform the analogue of the discussion of sec. \ref{sec:boson} to the lower-dimensional cases $d=3,2$ not captured by sec. \ref{sec:spinor}. This analysis is quite general, since it can covers both fermionic and bosonic representations simultaneously. 

\subsection{Kernel in $d=3$}

The highest dimension where our spinorial harmonic decomposition breaks down is $d=3$. The projectors over definite $\gamma\cdot p$ eigenvalue is no longer independent from the projectors $\Pi_{\pm \pm}$. To properly assess this case, we return to a formalism which is more dimension-specific. 

A general rotational representation $\rho$ of $Spin(3)$ is given by the spin-j representation $[j]$, with $j\in \frac{1}{2}\mathbb{N}$. This representation corresponds to states or fields $\mathcal{O}_{\alpha_1 \ldots \alpha_j}=\mathcal{O}_{(\alpha_1 \ldots \alpha_{2j})}$, sitting in the irreducible component of the tensor product of $2j$ spinors \cite{Iliesiu:2015qra,Iliesiu:2017nrv}. The fundamental spinor representation is isomorphic to $\mathbb{C}^2$, and the gamma matrices can be taken to be the Pauli matrices 
\Bal
	\sigma_1 &= \begin{pmatrix}
		0 & 1 \\ 1 & 0 
	\end{pmatrix} \, ,& \sigma_2 &= \begin{pmatrix}
		0 & i \\ -i & 0 
	\end{pmatrix} \, ,  & \sigma_3 &= \begin{pmatrix}
		1 & 0 \\ 0 & -1 
	\end{pmatrix} \, .
\Eal

We define index-free fields by contracting them with commuting polarisation spinors $\overline{s}^\alpha$, and $s_\alpha$. The Kernel must then reduce to the unique conformally invariant 2pt. function of these fields
\Bal
	G_{\Delta,j}(x,y)&=\expval{\mathcal{O}_{j}(x,\overline{s}_1)\overline{\mathcal{O}}_j(y,s_2)}=C_{\Delta,j}\frac{\left(i\, \overline{s}_1 \sigma\cdot(x-y) s_2 \right)^{2j}}{(x-y)^{2\Delta+2j}} \, ,
\Eal
Clearly, the polarisation of the correlator is encoded in the vector $z_\mu = \overline{s}_1 \sigma_\mu s_2$. Note that this vector is generically complex but not null. We can now look at the harmonic decomposition, the first step of which is to perform the Fourier transform
\begin{align}
	\widehat{G}_{\Delta,j}(p) = \frac{4^{h-\Delta-j} C_{\Delta,j} \pi^{h}\Gamma(h-\Delta-j)}{\Gamma(\Delta+j)}(-z\cdot \partial_p)^{2j} (p\cdot p)^{\Delta+j-h} \, .
\end{align}
The rightmost element can be related to an analytically continued Gegenbauer polynomial
\Bal
	(-z\cdot \partial_p)^{2j}(p\cdot p)^{-\alpha}&=\dv[j]{}{\epsilon} \left(p\cdot p -2 \epsilon z\cdot p + \epsilon^2 z\cdot z\right)^{-\alpha} \\
	& = (2j)!(p\cdot p)^{-\alpha}\left(\frac{z\cdot z}{p\cdot p}\right)^{j} C_{2j}^{\alpha}\left(t=\frac{z\cdot p}{\sqrt{z\cdot z p\cdot p}}\right) \, .
\Eal

the variable $t$ encodes the angle between the exchanged momenta and the spinorial polarisations,
\begin{align}
	z\cdot p = \sqrt{p\cdot p\  z\cdot z}\cos(\theta)= \sqrt{p\cdot p\  z\cdot z} \ t \, .
\end{align}

Since its index is negative, this Gegenbauer should properly be defined as a special hypergeometric function. With $\alpha = h- \Delta -j$, and $h=\frac{3}{2}$. For simplicity, we now set $\Delta = h + \nu$, giving 
\begin{align}
	C_{2j}^\alpha(t) = \frac{(-2\nu-2j)_{2j}}{(2j)!}{}_2F_1 \left(-2j,-2\nu ;\frac{1}{2}-\nu-j;\frac{1-t}{2}\right) \, .
\end{align}

\subsection{Harmonic Decomposition And $d=3$ Complementary Series}

To perform the harmonic decomposition, we will take a slightly different root as before, and use think of it more in term of an Hilbert space rather than a decomposition over special functions. Our states transform in (rotational) irreducible representations $[j]$ of $Spin(3)$, which is made of the span of the states $\ket{j,m}$, $2j \in \mathbb{N}$, $m = -j, -j+1 ,\ldots ,j$, following \cite{Kravchuk:2016qvl,Erramilli:2019njx}. These states are realised as the symmetrised tensor product of the fundamental, spin-$\frac{1}{2}$ irreducible representation. The stabiliser group of a given vector $p^{\mu}$ in $Spin(3)$ is given by a $Spin(2)$ subgroup, which irreducible representation $[m]$ is (complex) one-dimensional and transforms under rotation by angle $\phi$ as $\ket{m} \to e^{im\phi}\ket{m}$. The branching rule from $Spin(3) \to Spin(2)$ is very  simply 
\begin{align}
	\text{Res}^{Spin(3)}_{Spin(2)} [j]  = \sum_{m=-j}^{j}[m] \, .
\end{align}

By picking the basis of the fundamental representation to be the $\gamma\cdot p$ eigenvectors basis, this the branching rules become trivial, as each the states $\ket{j,m}$ directly encode the branching. Said otherwise, we can pick our basis choice so that each basis element transforms irreducibly under the stabiliser group. This means, to perform the harmonic decomposition, we simply have to look at the decomposition of the kernel over the finite dimensional basis.\footnote{One is free to chose some other decomposition procedure, for example Fourier series. This specific decomposition is simply the most physical one and makes the action of conformal symmetry the clearest.}
 
We can be more explicit now. We will label the basis states of the spinor representation $\ket{\frac{1}{2},\pm\frac{1}{2}} = \ket{\uparrow / \downarrow}$, which we take to be eigenstates of $\sigma\cdot p$. The polarisation spinor $s_\alpha$ then specifies a definite orientation of the spinors
\begin{align}\label{eq:decom1}
	s = \xi_\uparrow\ket{\uparrow}+\xi_\downarrow\ket{\downarrow} \, .
\end{align} 

Products of spinor polarisation define states living in the tensor product. The link to the irreducible spin-$j$ representations state is obtained through 
\begin{align}
	\ket{\uparrow}^{j+m}\ket{\downarrow}^{j-m}=i^{j+m}\binom{2j}{j+m}^{-\frac{1}{2}}\ket{j,m} \, .
\end{align}

Naturally, in this basis we have an equivalence between a certain collection of polarisation component and a projector
\Bal
	\overline{s}_1 A s_2 &=
	 \binom{2j}{j+m}\overline{\xi}_{1,\downarrow}^{j-m}\overline{\xi}_{1,\uparrow}^{j+m}\overbrace{\bra{j,m}A\ket{j,m}}^{A_{j,m}}\xi_{2,\downarrow}^{j-m}\xi_{2,\uparrow}^{j+m} \, ,  \\
	 A &= A_{j,m}\ket{j,m}\bra{j,m} \, .
\Eal
Which means that we can clear out our expressions, by ``freeing'' the indices, replacing the components of the polarisation spinor by projectors. Effectively, we take our previous expressions, evaluate them for the specific decomposition \eqref{eq:decom1}, and replace 
\begin{align}
	\overline{\xi_{1,a}}\xi_{2,b} \to  \ket{a}\bra{b} \, ,
\end{align} 
and we will understand expressions with multiplications of these kets to act with the tensor product, i.e. $(\ket{a}\bra{b})^2 = \ket{a}^2\bra{b}^2$. More prosaically, the kets $\ket{\uparrow/\downarrow}$ are to be viewed as shorthand for specific functions of the polarisation spinors with definite transformation properties.

Our goal is now to expand the kernel in term of projectors over the orthonormal basis vectors $\ket{j,m}$. An explicit computation gives
\Bal
	\sqrt{z\cdot z} &= \ket{\uparrow}\bra{\uparrow}+\ket{\downarrow}\bra{\downarrow}  \, ,  & z\cdot p &= \ket{\uparrow}\bra{\uparrow}-\ket{\downarrow}\bra{\downarrow} \, .
\Eal
The first result follows from a Fierz identity. We then use an hypergeometric identity to rewrite
\Bal
	\widehat{G}_{\Delta,j}(p) = &C_{\Delta,j}\frac{\pi ^h (-1)^{2j} 4^{j-\nu } \Gamma (j-\nu )}{\Gamma (\Delta+j)}(p\cdot p)^{\nu} (\ket{\downarrow}\bra{\downarrow})^{2j} \\
	&\times {}_2F_1\left(-2 j,\frac{1+2\nu-2j}{2};1+2\nu-2 j;1+\frac{\ket{\uparrow}\bra{\uparrow}}{\ket{\downarrow}\bra{\downarrow}}\right) \\
	=&(p\cdot p)^{\nu} N_{\Delta,j}\sum_{m=-j}^{j} \kappa_{j,m}(\Delta)  \ket{j,m}\bra{j,m}\label{eq:decomp2} \, .
\Eal
The harmonic coefficients can be extracted out of the $(j+m)$-th term in the series expansion of the function on the second line
\begin{align}
	\frac{1}{(j+m)!}\binom{2j}{j+m}^{-1} \dv[j+m]{}{x}{}  {}_2F_1\left(-2j,\frac{1+2\nu-2j}{2}; 1+2\nu-2j;1+x\right) \, ,
\end{align}
Which can be simplified as
\Bal 
	\kappa_{j,m}(\Delta) &=  (-1)^{j-m}\frac{ \Gamma \left(\Delta+m-1\right) \Gamma \left(\Delta-m-1\right)}{\Gamma \left(\Delta+j-1\right) \Gamma \left(\Delta-j-1\right)} \, , \\
	N_{\Delta,j} &= C_{\Delta,j} \frac{4 \pi  (\Delta -1) \Gamma (2-2 \Delta )}{\Delta +j-1}\sin (\pi  (\Delta +j)) \, .
\Eal
These match the result found in \cite{Karateev:2018oml,Erramilli:2019njx} using recursion relation, up to some global phase different due to the signature. These coefficients satisfy the relations
\begin{align}
	\kappa_{j,m}(\Delta)\kappa_{j,m}(d-\Delta) = 1  \, ,
\end{align}
Which imply the shadow transform indeed generates an automorphism, away from exceptional points. 

With these tools set, we can discuss whether the kernel $G_{\Delta,j}$ defines a unitary inner product. 

\paragraph{Normalisability :} The general discussion of the scalar case in sec. \ref{sec:boson} still applies, as it was dimension agnostic. The harmonic decomposition is positive provided $0<\Delta <d$. Beyond this, some modes can become non-normalisable and must be checked case by case, as in the scalar-exceptional series. 

\paragraph{Bosonic Representations :} We set $j \in \mathbb{N}^{\star}$. The harmonic coefficients must not switch signs for different values of m, which can be checked by requiring 
\begin{align}
	0\leq \frac{\kappa_{j,m+1}}{\kappa_{j,m}}=\frac{\Delta+m-1}{\overline{\Delta}+m-1}  \, .
\end{align}

The strongest bound comes from $m=0$, and gives $1<\Delta <d-1$, as previously. The kernel is irregular and admits unitary irreducible exceptional series representations for $\Delta = - n$, $n\in \mathbb{N}$ for $j=0$, and $\Delta = -t$, $0 \leq t<j$. Again, this reduces to the general discussion of \ref{sec:boson}.

\paragraph{Fermionic Representation :} We have $j \in  \frac{1}{2}+\mathbb{N}^{\star}$ and $m$ is an half-integer. The previous argument can be reproduced, and the strongest bound comes $m=-\frac{1}{2}$ which imposes $\Delta = \overline{\Delta}=\frac{3}{2}$, the tip of the principal series. In fact, opposite $m$ have contributions with identical norms but opposite signs
\begin{align}
	\frac{\kappa_{l,m}}{\kappa_{l,-m}} = (-1)^{2m} \, ,
\end{align}
meaning there is no room for complementary series fermions. This is really another avatar of the higher-dimensional pairing of eigenvalues. Given this situation, it is extremely surprising that one can construct unitary extraordinary fermionic representation. The procedure to do so, which we investigate in the next section, hinges on a decoupling of the modes with opposite $m$ at the exceptional points.

\subsection{Harmonic Gap And $d=3$ Exceptional Series}

The harmonic decomposition we have computed has the interesting feature that for specific values of the scaling dimension, $\Delta = d-1+j-r=2+j-r$ with $r=1,\ldots \lfloor j\rfloor $, the middle modes with $m\leq j-r$ decouple, creating an harmonic gap. The opposite happens for the conjugate dimensions $\overline{\Delta}=1-j+r$. With this gap, the modes with positive and negative $m$ become isolated, which allows us to split up the inner product 
\begin{align}
	G_{2+j-r,j} \to G^{+}_{2+j-r,j} \oplus G^{-}_{2+j-r,j} \, .
\end{align}

That this is a conformally invariant separation, is counter-intuitive : we have written the most general 2-point function and obtained a sum over both positive and negative helicities. In fact, one can check that under a special conformal transformation, the modes with different values of $m$ mix together into their nearest neighbors.  For the kernel written in the form \eqref{eq:decomp2}, conformal invariance is equivalent to imposing invariance under $K_\mu$, giving a recursion relation \cite{Karateev:2018oml} 
\begin{align}
	(\Delta+m-2)\kappa_{j,m}&=(\Delta-m-1)\kappa_{j,m-1} \, ,
\end{align}
that is solved by our previous solution.\footnote{In higher dimension we have seen that the up and down helicities mix together under special conformal transformation. This is the same thing happening here, the difference is that in higher dimension the harmonic decomposition could factor out the projection along $p\cdot \gamma$, and it acted as an overall factor. In $d=3$ meanwhile, this interacts with the rest of the helicities label, since $K_\mu$ mixes one of the $2j$ spinor substates with its opposite helicities, and  overall this relates states which differ by $m+1$,$m$ and $m-1$. The specific coefficients are then not even needed to understand that the harmonic gap decouples the two sectors. It is really the structure of the harmonic decomposition which changed with respect to higher dimensions.} What changes now, is that for the exceptional points, the harmonic gap opens up, isolating the two sets of modes. Hence, for this specific choice of dimensions, one can write a kernel improved by contact and improvement terms, whose harmonic decomposition only contains only one set of modes. 

For concreteness, consider the fully-conserved (bosonic or fermionic) field, i.e. $\Delta = j+1$, $r=1$. The modes $m=\pm j$ are totally isolated and do not mix under conformal transformations. One can write out a conformally covariant kernel which picks them out, 
\Bal
	\widehat{G}_{j+1,j}^{\pm}(p)&=(p\cdot p)^{j-\frac{1}{2}} N_{j+1,j}\bra{j,\pm j}\ket{j,\pm j}  \, , \\
	G_{j+1,j}^{\pm}(x,y)&=\frac{(i\,  \overline{s}_1 \sigma\cdot (x-y)s_2)^{2j}}{2\left(x-y \right)^{4j+2}}\pm \frac{\pi^2(\overline{s}_1 s_2) (i\, \overline{s}_1 \sigma\cdot\partial_x s_2)^{2j-1}}{(2i)^{2j}(2j-1)!} \delta^{3}(x-y) \, .
\Eal
Expressions for generic $r$ can be found case by case by an inverse Fourier transform. They take the form of a   sum of terms sprinkling in factors of $ \overline{s}_1 s_2$ and delta-functions. These improvement terms appear with an overall sign which picks out the positive or negative helicities. This feature, unique to $d=3$, has implications for both bosonic and fermionic representations. 

\paragraph{Bosonic representations :} The analysis carried out in sec. \ref{sec:boson} still applies, and the representation $\mathcal{U}_{\ell,t}$ is unitary and can be identified with the one extracted out of the previous discussion under the identification $t=\ell-r$. This representation is however not irreducible anymore, since it has a conformally invariant subspace made up of the positive and negative helicities. These two subspaces are conjugate under reflections
\begin{align}
	\mathcal{U}_{j,t}\underset{d=3}{\longrightarrow} \mathcal{U}_{j,t}^{+} \oplus \mathcal{U}_{j,t}^{-} \, . 
\end{align}

The two ``chiral'' representations, $\mathcal{U}_{\ell,t}^{\pm}$, are the unitary irreducible representations relevant for 3-dimensional partially conserved tensors or depth $j-t$. These representations happen to be discrete series, meaning their group element are square integrable \cite{Dobrev:1977qv}. In a parity invariant theory, one will obtain both of these representations in pairs and the relevant inner-product will possess no improvement terms.

\paragraph{Fermionic representation :} As we have seen, the fermions are plagued by opposite mode having opposite signs. For the exceptional points at  $\Delta = d-1+j+k$, $k\in \mathbb{N}$, there is no harmonic gap and this problem persist. Even worse, one finds that modes on both sides alternate in signs. The only hope is for the previously mentioned representations with $\Delta = d-1+j-r$. There, because of the gap, we can consistently flip the sign of the kernel for the positive and negative helicity modes so that both give rise to positive inner product : $G^{+}_{2+j-r,j}$ and $-G^{-}_{2+j-r,j}$ are both positive-definite, and both give rise a different representation $\mathcal{U}_{j,t}^{\pm}$, which are a conjugate under reflections. Their sum is a positive, conformally invariant inner-product on the fermionic representation, in which only the improvement terms survive . It is furthermore clear that such a representation breaks parity explicitly. This was evident already in the treatment of \cite{Letsios:2023qzq,Letsios:2023awz,Letsios:2024nmf} which had found a positive inner-product over the basis of mode functions which involved $\gamma_\star$.\footnote{From a $d=3$ perspective, it is not very surprising that (exceptional) fermions break parity. For example, the mass term for a free $d=3$ fermion breaks parity as well.} These representations are the exceptional partially-conserved higher-spinors with depth $\lfloor j-r\rfloor$.

\subsection{The Complex Plane And $d=2$}

We now investigate the case of $d=2$. In what follows, $h$ will be a generic complex weight and not $d/2=1$. Famously, the spin-group is isomorphic to the Möbius group $Spin(3,1) \simeq SL(2,\mathbb{C})$ \cite{Bargmann:1946me,Gelfand:1947aa,Duc:1967aa,Bonifacio:2023ban} . Owing to the decomposition $SL(2,\mathbb{C})\sim SL(2,\mathbb{R})\times SL(2,\mathbb{R})$, we work with states and wavefunctions in complex coordinates, with no holomorphicity assumptions. These are totally equivalent to complex states and wavefunctions of $2$-real variables under the identification
\begin{align}
	(x,y) \to z=x+iy \, .
\end{align}
Induced representations are those which transform with weight $(h,\overline{h})$ under the $SL(2,\mathbb{R})\times SL(2,\mathbb{R})$ subgroup, which is manifest in the $(z,\overline{z})$ coordinates \cite{DiFrancesco:1997nk},
\Bal
	L_{0}\ket{h,\overline{h};0}&=h\ket{h,\overline{h};0} \, , & \overline{L}_{0}\ket{h;\overline{h};0}&=\overline{h}\ket{h,\overline{h};0}  \, ,\\
	L_{-}\ket{h,\overline{h};0}&=0  \, ,& \overline{L}_{-}\ket{h,\overline{h};0}&=0 \, ,
\Eal
and the states at generic coordinate $(z,\overline{z})$ is constructed through 
\begin{align}
	\ket{h,\overline{h};z}=e^{zL_{+}+\overline{z}\overline{L}_{+}}\ket{h,\overline{h};0} \, .
\end{align}
Having specified the transformation laws, we can define the smeared states \cite{Bonifacio:2023ban}. We pick as a measure over the complex number the usual Lebesgue measure over $\mathbb{R}^2$ 
\begin{align}
	\int_{\mathbb{C}} \D^2z= \frac{i}{2}\int_{\mathbb{C}}\D z\wedge \D \overline{z}=\int_{\mathbb{R}^2} \D\Re(z) \, \D\Im(z) \, .
\end{align}
This measure is invariant under $z \leftrightarrow \overline{z}$, as is clear from writing it in the real parametrisation. This complex-space realisation can be slightly cumbersome at times, because one must be careful about how complex-conjugation act on the function versus the coordinates. For us, complex conjugation will always act on the function, leaving the coordinates unchanged. Everything is now setup to define our smeared states
\begin{align}
	\ket{\psi} = \int_{\mathbb{C}} \D^{2}z \  \psi(z)\ket{z;h,\overline{h}} \, .
\end{align}
By going back to the real coordinates, we have the identification $h+\overline{h}=\Delta$ and $h-\overline{h}=m$. Single-valuedness implies $2m\in \mathbb{N}$. Such a state or wavefunction transforms in the (one-dimensional) representation $[m]$ of the $Spin(2)\equiv U(1)$ subgroup. Finally, in the complex variables, the asymptotic condition on the wavefunctions becomes
\begin{align}
	\lim_{\abs{z}\to \infty}\psi_{h,\overline{h}}(z,\overline{z})\approx z^{h-1}\overline{z}^{\overline{h}-1} \, .
\end{align}

These requirements define the elementary representation $\mathcal{F}_{2-\Delta,m}$ appropriate to $Spin(3,1)$. We equip this space with an inner product
\begin{align}
	\bra{f_1}\ket{f_2} = \int_{\mathbb{C}^2}\D^2x\ \D^2y \ \overline{f_1(x)}f_2(y)G_{h,\overline{h}}(x,y) \, ,
\end{align}
that is sesquilinear and hermetian. The kernel can be identified as the 2pt. function of quasi-primaries fields in an Euclidean 2D CFT
\begin{align}
	G_{h,\overline{h}}(x,y) &= \frac{i^{2m}C_{h,\overline{h}}}{(x-y)^{2h}(\overline{x}-\overline{y})^{2\overline{h}}} \, ,
\end{align}
Which under simultaneous complex conjugation and exchange $x\leftrightarrow y$ is left invariant. This ensures that $\bra{f_1}\ket{f_2}^{\dagger} = \bra{f_2}\ket{f_1}$. 

To check for unitarity, we must decompose this kernel over a distinguished basis. We can again rewrite the inner product in Fourier space, using the complex-version of the Fourier transform
\Bal
	\widehat{f}(p) &= \int_{\mathbb{C}}\D^2z\,  e^{\overline{z}p-z\overline{p}}\, f(z) \, , & f(z) &= \int_{\mathbb{C}}\frac{\D^2p}{\pi^2}\,  e^{\overline{p}z-p\overline{z}}\,\widehat{f}(p)  \, , \\ 
	\delta^2(z)& = \int_{\mathbb{C}} \frac{\D^2p}{\pi^2}\, e^{\overline{p}z-p\overline{z}} \, , &
	\left(\widehat{f}(p)\right)^{\dagger} &= \int_{\mathbb{C}}\D^2z\,  e^{\overline{z}p-z\overline{p}}\, f^{\dagger}(z) = \widehat{f}^{\dagger}(-p) \, .
\Eal
These types of relations and transform are sometimes called Wigner-Weyl transform, and are used in Quantum optics in the study of coherent states, see for example \cite{Gerry_Knight_2004}. The inner product becomes 
\begin{align}
	\bra{f_1}\ket{f_2} =\int_{\mathbb{C}} \frac{\D^2p}{\pi^2} \, \left(\widehat{f_1}(p)\right)^{\dagger} \, \widehat{f_2}(p)\, \widehat{G}_{1-\overline{h},1-h}(p) \, .
\end{align}

Having set up these conventions, we go forward with the Fourier transform, 
\begin{align}
	\widehat{G}(p)&=i^{2m}C_{h,\overline{h}}\int_{0}^{\infty}\frac{\D r}{r^{2\Delta-1}} \int_{0}^{2\pi} \D\varphi \  e^{2i\abs{p} r cos(\varphi-\arg(p))-2 i m \varphi} \, .
\end{align}
The exponential can be rewritten as an infinite sum of Bessel functions,

\Bal
	\hat{G}_{h,\overline{h}}(p) &= i^{2m} C_{h,\overline{h}}\sum_{n=-\infty}^{\infty} \left(-ie^{-i\arg(p)}\right)^{n}\int_{0}^{\infty}\D r\, \frac{J_{n}(2\abs{p}r)}{r^{2\Delta-1}} \int_0^{2\pi} \frac{\D \phi}{(2\pi)^2} \, e^{i(n-2m)\phi} \\
	&=p^{2\overline{h}-1}\ \overline{p}^{2h-1}\ \frac{C_{h,\overline{h}}}{2\pi}\frac{\Gamma \left(1+m-\Delta\right)}{\Gamma \left(m+\Delta\right)} \, .
\Eal

This kernel as quite a simple form, since we obtain two overall $p^{2\Delta-d}$ with $\Delta \sim h,\overline{h}$ and $d\sim 1$, as we would expect for a 1-dimensional Fourier transform of an $SL(2,\mathbb{R})$ 2pt. function. This kernel is normalisable for $0<\Delta<2$, as it is expected from the scalar analysis. To investigate the positivity, let us pick $\bra{f}\ket{f}$, which is real, and expand the $\widehat{f}(p)$ in Fourier modes over the argument of $p$
\Bal
	\bra{f}\ket{f}&\propto \sum_{n,k}
	\int_0^{\infty} \D r\, r^{2\overline{\Delta}-1} \int_{-\pi}^{\pi} \D\phi\, e^{i(n-k-2m)\phi} \, f_n(r)f_k^{\dagger}(r) \\
	&= \sum_{n} \int_{0}^{\infty} \D r \, r^{2\overline{\Delta}-1}  \Re\left(f_{n}(r)f_{n+2m}^{\dagger}(r)\right) \, .
\Eal
This final sum is not over positive numbers, except if $m=0$. This implies that it is impossible to generate non-trivial spinning exceptional representations, since those would always have negative norm sectors. This can be understood differently, by noting that the subspace $\mathcal{U}_{\ell,t}$ is zero-dimensional, hence we are only left with non-unitary invariant subspaces at exceptional points.

We have established that the unitary irreducible representations are the principal series $\mathcal{P}_{1+i\nu,m}$, and the scalar complementary series $\mathcal{C}_{\Delta}$, for $\Delta \in [0,2]$. We can consistently restrict ourselves to this half of the segment since the other half is equivalent under the shadow transform.

\begin{figure}[H]
	\centering
	\includegraphics[width=0.45\linewidth]{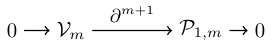}
	\caption{In $2D$, the scalar exceptional representation is isomorphic to the tip of the spinning principal series.}
	\label{fig:redundant}
\end{figure}

It is very reasonable at this point to wonder about the exceptional representations, which were so important in other dimensions. They turn out not to be new representations in $d=2$. First, consider representations with non-zero spin $m$, they clearly suffer from the same non-positivity as the complementary spinning representation, and so are ruled out. Consider now the scalar exceptional series, as constructed in sec. \ref{sec:exscal}. The previous logic does go through unchanged, and one can build these representations, but they are redundant with the spinning principal series.\footnote{This is actually a feature which also appears in higher dimensions. One can replace an exceptional representation by a principal series one with more involved spin structure. Exceptional representation are defined as equivalence classes, i.e. they have a gauge symmetry. From a field perspective, the logic is to build an invariant field-strength that picks out a given equivalence class. See comments in \cite{Dobrev:1977qv}.} The reasoning is summarised in fig. \ref{fig:redundant}, which we now spell out. 
\noindent Consider the inner product for $\mathcal{F}_{-m,0}$, $m\in \mathbb{N}$, which gives non-zero norm to the states isomorphic to $\mathcal{V}_m$. It is generated by the kernel 
\begin{align}
	G_{1+\frac{m}{2},1+\frac{m}{2}}&\propto\frac{1}{(x-y)^{1+m}(\overline{x}-\overline{y})^{1+m}} \leadsto (-\partial_x\overline{\partial}_x)^{1+m}\delta^{2}(x-y)  
\end{align}
which is effectively ultra-local because of Cauchy's theorem.\footnote{This can be seen as an instance of the general non-conformality previously discussed, for which the proper conformally invariant 2pt. function with scaling $\Delta = \frac{d}{2}+k$ is the ultra-local one like here. } The inner product of $f_1,f_2 \in \mathcal{V}_m \in \mathcal{F}_{-m,0}$ is then 
\begin{align}
	\bra{f_1}\ket{f_2} = \int \D^{2}z \, \left( \partial^{m+1} f_1(z) \right)^{\dagger} \partial^{m+1}f_2(z) = \left( \partial^{m+1} f_1, \partial^{m+1} f_2 \right) \, ,
\end{align}
where $\partial$ and $\overline{\partial}$ play the role of $d$ and $\overline{d}$ in higher dimension as weight-shifting operators taking us from one representation to another. Since they increase the weight $h$, respectively $\overline{h}$, by $1$, we find that $\partial^{m+1}f \in \mathcal{P}_{1,1+m}$. Since all of $\mathcal{V}_m$ is outside of the kernel of $G_{1+\frac{m}{2},1+\frac{m}{2}}$ by construction, this concludes showing the equivalence between the representations.

One can re-understand this construction in light of the usual CFT formalism for the $2D$ free boson. The free boson $\phi$ has $\Delta = 0$, which correspond to $m=0$. Its correlation functions are logarithmic, as predicted by the fact that it sits in an exceptional representation and is invariant under constant shifts $\phi \to \phi + c$. One can however construct a primary operator $\partial\phi$ which is shift-invariant, and which sits in the principal series representation with $\Delta = 1$. The output of this discussion is that for euclidean conformal field theories, one has more unitary representations of this type to consider, which are generalisations of this construction with more derivatives.

\section{Discussion}

We have shown how to construct a few families of generic unitary irreducible representations of the de Sitter group, the symmetric traceless tensor and tensor-sinors. This extends the current literature, giving a novel account of these representations, which are most relevant to the study of fermionic operators in expanding space. The two important outputs of this work are the following. First, our discussion of exceptional representations stresses the great departure from the usual rules of conformal field theories as encountered in Lorentzian signature. Exceptional representations have logarithmic correlators, which far from breaking de Sitter isometries, are mandated by them. Second, we have shown and explained the structure of fermionic representations. We explained why spinor fields only exist in the principal series, except in $d=3$, which allows remarkably for the existence of partially-massless unitary representations because. This boils down to the fact that the spinor representation possesses a pair of opposite helicities for the general case, and to a degeneracy of the Clifford algebra in $d=3$ for the exceptional one. 

The question of the field theoretic realisation of these unitary representations remains an open problem, specifically given the naive complex-valued Lagrangian description of massless higher spinors \cite{Pilch:1984aw,Letsios:2023qzq,Bonifacio:2023prb}. This is part of a broader problem regarding exceptional representations in de Sitter. The question for $d=1$ discrete series scalars has only recently been resolved \cite{Anninos:2023lin}. There is moreover evidence for an analogue of Weinberg-Witten theorem \cite{Weinberg:1980kq} for partially-massless tensors \cite{Sleight:2021iix}, hinting at a rich structure to be explored in the future.

As already acknowledged in the introduction, there are plenty of future directions to be pursued. Notably, given the construction of fermionic representations given, it would be extremely useful to use it to extend our knowledge of multi-particle states in de Sitter following \cite{Dobrev:1976vr,Penedones:2023uqc}. For example, it would be very interesting to see whether the tensor-product of principal series spinor overlaps with bosonic exceptional representations, and whether the tensor-product of bosons and fermions can overlap with the exceptional spinors in $dS_4$. These considerations are essential to the full development of the bootstrap approach to cosmological correlators \cite{Hogervorst:2021uvp}. A related issue is the study of the characters, \cite{Hirai:1965aa,Harish-Chandra:1966aa} and their link with quasinormal modes \cite{Du:2004jt,Lopez-Ortega:2006aal,Lopez-Ortega:2012xvr,Jafferis:2013qia,Sun:2020sgn,Law:2023ohq}. This would help gain more understanding of fermionic contributions to the path integral in de Sitter in generic dimensions, \cite{Anninos:2020hfj,Law:2020cpj,Muhlmann:2021aa,Law:2023ohq}, which hopefully would then build further into the investigation of the properties of the cosmological horizon and the dynamic in the static-patch  \cite{Anninos:2019aa,Albrychiewicz:2020ruh,Anninos2021,Harris2021,Mirbabayi:2022gnl}.

\acknowledgments

The author would like to thank V.~Letsios for first bringing the partially-massless spinor to his attention and useful discussion regarding his work, D.~Anninos, T.~Anous,, A.~Rios-Fukelman and B.~Pethybridge for discussions regarding the discrete series representations, and P.~Kravchuk for his insights about the conformal group and field theory. Thanks are given to  C.~Herzog, D.~Anninos for their comments on the drafts and constant encouragements. The author is grateful to the organisers of the workshop ``Holography in \& Beyond the AdS Paradigm'' at Instituto de Fisica de La Plata which overlapped with the elaboration of this work, and to M.~Downing, G.~Watts, for useful discussions about $2D$ CFT. This work was supported by the U.K.\ Science \& Technology Facilities Council Grant ST/P000258/1.

\appendix 

\section{Discrete Versus Exceptional And Parity of $d$}
\label{app:discrete}

The unitary representation split between representations which are regular (principal and complementary series), on which the shadow transform defines an isomorphism, and the exceptional series. The exceptional representations have many interesting features : their inner product is logarithmic, naively breaking the isometries, their Hilbert space has a quotient structure, they are non-trivially unitary, etc. The analysis of the unitary representations we performed is fully general, and include as a sub-case the relevant discrete series representations. 

The discrete series form a subset of the exceptional representations, for which the group-element are square integrable over the group manifold  \cite{Harish-Chandra:1965aa,Harish-Chandra:1966aa,Dobrev:1977qv}. More concretely, for $g \in G = SO(1,d+1)$, and $\psi$ transforming in some representation $\mathcal{R}$, define $g\circ \psi = R(g)[\psi]$. $R(g)$ is a functional which is the realisation of $g$ on the representation $\mathcal{R}$. Such a representation is said to be discrete if
\begin{align}
	\int_{G} 
 \abs{\expval{\psi_1,R(g)\psi_2}}^2  \D g   &=\frac{\abs{\expval{\psi_1,\psi_1}\expval{\psi_2,\psi_2}}^2}{d_R} \, , \quad  d_R>0  \, .
 \end{align}
Where $\D g$ is the Haar measure over the group $G$. Peeling back the complexity, picking an orthonormal basis of the representation we see this is simply the statement that the group-average norm of any elements of the representation matrices are equal, proportional to the identity, and finite. $d_R$ is called the formal dimension of the representation. For most representations of non-compact Lie groups, this integral diverge, and there is no such non-zero $d_R$.

This normalisability, is highly dimensionally dependant, and totally unrelated to the normalisability requirement for the element of the unitary representations. This gives rise to a peculiar picture which can offer a different lighting on some of the results of this work. We would like to stress however, that the fact that a given exceptional representation is \textit{discrete} does not seem to bring  special properties outside of concerns regarding group-averaging.

To understand the pattern of the discrete series, one must (re)-consider the size of the group-factors. The general statement is due to Harish-Chandra \cite{Harish-Chandra:1965aa}, and is that discrete series representations exist when the maximal compact subgroup has the same rank as the group. Our representations are constructed in the the non-compact picture from the factors $(A,M)$, with $M=SO(d)$. let us call $H_{G}$ the Cartan subalgebra of a group $G$. Our induced representations are then constructed of the full Cartan  $A \oplus H_M$, which is non-compact and has dimension $1+\text{dim}(H_{M})$. The non-compactness of the Cartan translates into the generic non-normalisability of the group elements over the group manifold.

Consider now the compact picture, where one uses the Iwasawa decomposition $KAN$, with $K=SO(d+1)$. If $d=2k$, there are the Lie-algebra identifications $M \equiv D_k$ and $K\equiv C_k$, and both have the same rank, and the non-compact Cartan subalgebra has dimension $k+1$. However, if $d=2k+1$, then $M \equiv C_K$ and $K \equiv D_{k+1}$. Then, the Cartan has dimension $k+1$, and it can be picked to be the compact $H_{K}$. This gives rise to integrable representations, which are the discrete series.
\Bal
	SO(2k) &= D_k  \quad \includegraphics[valign=c,width=0.2\linewidth]{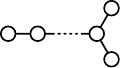}  \, , \\
	SO(2k+1) &= B_k  \quad \includegraphics[valign=c,width=0.25\linewidth]{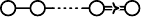} \, .
\Eal

These group theoretical considerations explain the pattern, though it is far removed from the more physical intertwining kernel perspective, which is what actually builds up the representation. From the point of view of the Hilbert space construction, nothing drastic happens from one dimension to the next. Take for example the spinning exceptional series $\mathcal{U}_{\ell,t}$. For $d=3$, these happen to break in two pieces, which are discrete series. This is also the case for the scalar exceptional series $\mathcal{V}_{n}$ for $d=1$. None of the interesting physical properties of these representations is due to the discreteness per se, since these features survive, unchanged, in arbitrary dimensions. 

Part of the features of fermionic representations can be understood through this lense. The partially massless tensor-spinors, the would-be fermionic equivalents of $\mathcal{U}_{\ell,t}$, are non-unitary in all dimensions, except when they fall in the discrete series and this only happens in $d=3$. In $d=1$,
a similar process occurs for the fermionic equivalents of $\mathcal{V}_{n}$ \cite{Kitaev:2017hnr,Sun:2021thf}. The fact that these representations are not unitary for other dimension is however much better understood as a feature of inner-product on the representation, rather than some rather global, topological statement regarding the representation of group elements.

\section{Reminder About Exact Sequences}\label{app:seq}

In this work, we sometimes encounter exact sequences of a particular type, where all spaces are vector spaces and all the maps are linear. These are all split exact sequences. For convenience, we here collect some of the basic properties we use throughout. 

\begin{figure}[H]
	\centering
	\includegraphics[width=0.45\linewidth]{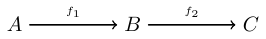}
	\caption{Typical exact sequence, with $\Im(f_1)=\text{Ker}(f_2)$. In this work, all maps will be linear, between vector spaces, making these split sequences.}
	\label{fig:seq1}
\end{figure}

\begin{figure}[H]
\centering
\begin{subfigure}{0.45\textwidth}
	\centering
    \includegraphics[width=0.7\textwidth]{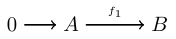}
    \caption{$f_1$ is injective.}
\end{subfigure}
\centering
\begin{subfigure}{0.45\textwidth}
	\centering
    \includegraphics[width=0.7\textwidth]{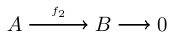}
    \caption{$f_2$ is surjective.}
    \label{fig:second}
\end{subfigure}
\hfill
\begin{subfigure}{0.8\textwidth}
	\centering
    \includegraphics[width=0.5\textwidth]{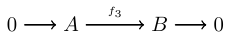}
    \caption{$f_3$ is a bijective map which defines an isomorphism $A \simeq B$.}
    \label{fig:third}
\end{subfigure}
        
\caption{Examples of properties encoded in exact sequences. We usually omit the leftmost $0$ for injective maps. In particular, for $A \subset B$, we only write the inclusion map $\iota$ from $A$ to $B$.}
\label{fig:seq2}
\end{figure}

An exact sequence is a diagram where spaces are connected by series of map with definite properties, as summarised in fig. \ref{fig:seq1}.
In complicated diagrams, the rules are that going two steps gives zero. If the endpoints are empty sets, then the maps have further properties, illustrated in fig. \ref{fig:seq2}

\bibliographystyle{jhep}
\bibliography{paper.bib}

\providecommand{\href}[2]{#2}\begingroup\raggedright\begin{thebibliography}{100}

\bibitem{Wigner:1939cj}
E.P.~Wigner, \emph{{On Unitary Representations of the Inhomogeneous Lorentz
  Group}}, \href{https://doi.org/10.2307/1968551}{\emph{Annals Math.}
  {\bfseries 40} (1939) 149}.

\bibitem{Weinberg:1995mt}
S.~Weinberg, \emph{{The Quantum theory of fields. Vol. 1: Foundations}},
  Cambridge University Press (6, 2005).

\bibitem{Hubble:1929ig}
E.~Hubble, \emph{{A relation between distance and radial velocity among
  extra-galactic nebulae}},
  \href{https://doi.org/10.1073/pnas.15.3.168}{\emph{Proc. Nat. Acad. Sci.}
  {\bfseries 15} (1929) 168}.

\bibitem{Bousso:2007gp}
R.~Bousso, \emph{{TASI Lectures on the Cosmological Constant}},
  \href{https://doi.org/10.1007/s10714-007-0557-5}{\emph{Gen. Rel. Grav.}
  {\bfseries 40} (2008) 607} [\href{https://arxiv.org/abs/0708.4231}{{\ttfamily
  0708.4231}}].

\bibitem{Baumann:2020ksv}
D.~Baumann, C.~Duaso~Pueyo and A.~Joyce, \emph{{Bootstrapping Cosmological
  Correlations}},
  \href{https://doi.org/10.22661/AAPPSBL.2020.30.6.02}{\emph{AAPPS Bull.}
  {\bfseries 30} (2020) 2}.

\bibitem{Baumann:2020dch}
D.~Baumann, C.~Duaso~Pueyo, A.~Joyce, H.~Lee and G.L.~Pimentel, \emph{{The
  Cosmological Bootstrap: Spinning Correlators from Symmetries and
  Factorization}},  \href{https://arxiv.org/abs/2005.04234}{{\ttfamily
  2005.04234}}.

\bibitem{Baumann:2019oyu}
D.~Baumann, C.~Duaso~Pueyo, A.~Joyce, H.~Lee and G.L.~Pimentel, \emph{{The
  cosmological bootstrap: weight-shifting operators and scalar seeds}},
  \href{https://doi.org/10.1007/JHEP12(2020)204}{\emph{JHEP} {\bfseries 12}
  (2020) 204} [\href{https://arxiv.org/abs/1910.14051}{{\ttfamily
  1910.14051}}].

\bibitem{Baumann2018}
D.~Baumann, G.~Goon, H.~Lee and G.L.~Pimentel, \emph{Partially massless fields
  during inflation},
  \href{https://doi.org/10.1007/JHEP04(2018)140}{\emph{Journal of High Energy
  Physics 2018 2018:4} {\bfseries 2018} (2018) 1}.

\bibitem{Bros:1990cu}
J.~Bros, \emph{{Complexified de Sitter space: Analytic causal kernels and
  Kallen-Lehmann type representation}},
  \href{https://doi.org/10.1016/0920-5632(91)90119-Y}{\emph{Nucl. Phys. B Proc.
  Suppl.} {\bfseries 18} (1991) 22}.

\bibitem{Weinberg2005}
S.~Weinberg, \emph{Quantum contributions to cosmological correlations},
  \href{https://doi.org/10.1103/PHYSREVD.72.043514}{\emph{Physical Review D -
  Particles, Fields, Gravitation and Cosmology} {\bfseries 72} (2005) 1}.

\bibitem{Bros:2010rku}
J.~Bros, H.~Epstein and U.~Moschella, \emph{{Particle decays and stability on
  the de Sitter universe}},
  \href{https://doi.org/10.1007/s00023-010-0042-7}{\emph{Annales Henri
  Poincare} {\bfseries 11} (2010) 611}
  [\href{https://arxiv.org/abs/0812.3513}{{\ttfamily 0812.3513}}].

\bibitem{Sleight:2019mgd}
C.~Sleight, \emph{{A Mellin Space Approach to Cosmological Correlators}},
  \href{https://doi.org/10.1007/JHEP01(2020)090}{\emph{JHEP} {\bfseries 01}
  (2020) 090} [\href{https://arxiv.org/abs/1906.12302}{{\ttfamily
  1906.12302}}].

\bibitem{Sleight:2019hfp}
C.~Sleight and M.~Taronna, \emph{{Bootstrapping Inflationary Correlators in
  Mellin Space}}, \href{https://doi.org/10.1007/JHEP02(2020)098}{\emph{JHEP}
  {\bfseries 02} (2020) 098}
  [\href{https://arxiv.org/abs/1907.01143}{{\ttfamily 1907.01143}}].

\bibitem{Gorbenko:2019rza}
V.~Gorbenko and L.~Senatore, \emph{{$\lambda \phi^4$ in dS}},
  \href{https://arxiv.org/abs/1911.00022}{{\ttfamily 1911.00022}}.

\bibitem{Sengor:2019mbz}
G.~Seng\"or and C.~Skordis, \emph{{Unitarity at the Late time Boundary of de
  Sitter}}, \href{https://doi.org/10.1007/JHEP06(2020)041}{\emph{JHEP}
  {\bfseries 06} (2020) 041}
  [\href{https://arxiv.org/abs/1912.09885}{{\ttfamily 1912.09885}}].

\bibitem{Sleight:2020obc}
C.~Sleight and M.~Taronna, \emph{{From AdS to dS Exchanges: Spectral
  Representation, Mellin Amplitudes and Crossing}},
  \href{https://arxiv.org/abs/2007.09993}{{\ttfamily 2007.09993}}.

\bibitem{Sengor:2021zlc}
G.~Sengor and C.~Skordis, \emph{{Scalar two-point functions at the late-time
  boundary of de Sitter}},  \href{https://arxiv.org/abs/2110.01635}{{\ttfamily
  2110.01635}}.

\bibitem{Sleight:2021iix}
C.~Sleight and M.~Taronna, \emph{{On the consistency of (partially-)massless
  matter couplings in de Sitter space}},
  \href{https://doi.org/10.1007/JHEP10(2021)156}{\emph{JHEP} {\bfseries 10}
  (2021) 156} [\href{https://arxiv.org/abs/2106.00366}{{\ttfamily
  2106.00366}}].

\bibitem{DiPietro:2021sjt}
L.~Di~Pietro, V.~Gorbenko and S.~Komatsu, \emph{{Analyticity and Unitarity for
  Cosmological Correlators}},
  \href{https://arxiv.org/abs/2108.01695}{{\ttfamily 2108.01695}}.

\bibitem{Hogervorst:2021uvp}
M.~Hogervorst, J.a.~Penedones and K.S.~Vaziri, \emph{{Towards the
  non-perturbative cosmological bootstrap}},
  \href{https://doi.org/10.1007/JHEP02(2023)162}{\emph{JHEP} {\bfseries 02}
  (2023) 162} [\href{https://arxiv.org/abs/2107.13871}{{\ttfamily
  2107.13871}}].

\bibitem{Goodhew:2021oqg}
H.~Goodhew, S.~Jazayeri, M.H.~Gordon~Lee and E.~Pajer, \emph{{Cutting
  cosmological correlators}},
  \href{https://doi.org/10.1088/1475-7516/2021/08/003}{\emph{JCAP} {\bfseries
  08} (2021) 003} [\href{https://arxiv.org/abs/2104.06587}{{\ttfamily
  2104.06587}}].

\bibitem{Melville:2021lst}
S.~Melville and E.~Pajer, \emph{{Cosmological Cutting Rules}},
  \href{https://doi.org/10.1007/JHEP05(2021)249}{\emph{JHEP} {\bfseries 05}
  (2021) 249} [\href{https://arxiv.org/abs/2103.09832}{{\ttfamily
  2103.09832}}].

\bibitem{Heckelbacher:2022hbq}
T.~Heckelbacher, I.~Sachs, E.~Skvortsov and P.~Vanhove, \emph{{Analytical
  evaluation of cosmological correlation functions}},
  \href{https://doi.org/10.1007/JHEP08(2022)139}{\emph{JHEP} {\bfseries 08}
  (2022) 139} [\href{https://arxiv.org/abs/2204.07217}{{\ttfamily
  2204.07217}}].

\bibitem{Schaub:2023scu}
V.~Schaub, \emph{{Spinors in (Anti-)de Sitter Space}},
  \href{https://doi.org/10.1007/JHEP09(2023)142}{\emph{JHEP} {\bfseries 09}
  (2023) 142} [\href{https://arxiv.org/abs/2302.08535}{{\ttfamily
  2302.08535}}].

\bibitem{DiPietro:2023inn}
L.~Di~Pietro, V.~Gorbenko and S.~Komatsu, \emph{{Cosmological Correlators at
  Finite Coupling}},  \href{https://arxiv.org/abs/2312.17195}{{\ttfamily
  2312.17195}}.

\bibitem{Anninos:2024fty}
D.~Anninos, T.~Anous and A.~Rios~Fukelman, \emph{{de Sitter at all loops: the
  story of the Schwinger model}},
  \href{https://arxiv.org/abs/2403.16166}{{\ttfamily 2403.16166}}.

\bibitem{Loparco:2023rug}
M.~Loparco, J.~Penedones, K.~Salehi~Vaziri and Z.~Sun, \emph{{The
  K\"all\'en-Lehmann representation in de Sitter spacetime}},
  \href{https://arxiv.org/abs/2306.00090}{{\ttfamily 2306.00090}}.

\bibitem{Loparco:2023akg}
M.~Loparco, J.~Qiao and Z.~Sun, \emph{{A radial variable for de Sitter
  two-point functions}},  \href{https://arxiv.org/abs/2310.15944}{{\ttfamily
  2310.15944}}.

\bibitem{Galante:2023uyf}
D.A.~Galante, \emph{{Modave lectures on de Sitter space \& holography}},
  \href{https://doi.org/10.22323/1.435.0003}{\emph{PoS} {\bfseries Modave2022}
  (2023) 003} [\href{https://arxiv.org/abs/2306.10141}{{\ttfamily
  2306.10141}}].

\bibitem{Loparco:2024ibp}
M.~Loparco, \emph{{RG flows in de Sitter: c-functions and sum rules}},
  \href{https://arxiv.org/abs/2404.03739}{{\ttfamily 2404.03739}}.

\bibitem{Anninos:2014lwa}
D.~Anninos, T.~Anous, D.Z.~Freedman and G.~Konstantinidis, \emph{{Late-time
  Structure of the Bunch-Davies De Sitter Wavefunction}},
  \href{https://doi.org/10.1088/1475-7516/2015/11/048}{\emph{JCAP} {\bfseries
  11} (2015) 048} [\href{https://arxiv.org/abs/1406.5490}{{\ttfamily
  1406.5490}}].

\bibitem{Goodhew:2020hob}
H.~Goodhew, S.~Jazayeri and E.~Pajer, \emph{{The Cosmological Optical
  Theorem}}, \href{https://doi.org/10.1088/1475-7516/2021/04/021}{\emph{JCAP}
  {\bfseries 04} (2021) 021}
  [\href{https://arxiv.org/abs/2009.02898}{{\ttfamily 2009.02898}}].

\bibitem{Salcedo:2022aal}
S.A.~Salcedo, M.H.G.~Lee, S.~Melville and E.~Pajer, \emph{{The Analytic
  Wavefunction}},  \href{https://arxiv.org/abs/2212.08009}{{\ttfamily
  2212.08009}}.

\bibitem{Anninos2021}
D.~Anninos, T.~Bautista and B.~M{\"u}hlmann, \emph{The two-sphere partition
  function in two-dimensional quantum gravity},
  \href{https://doi.org/10.1007/JHEP09(2021)116}{\emph{JHEP} {\bfseries 09}
  (2021) 116}.

\bibitem{Strominger:2001pn}
A.~Strominger, \emph{{The dS / CFT correspondence}},
  \href{https://doi.org/10.1088/1126-6708/2001/10/034}{\emph{JHEP} {\bfseries
  10} (2001) 034} [\href{https://arxiv.org/abs/hep-th/0106113}{{\ttfamily
  hep-th/0106113}}].

\bibitem{Deser:2003gw}
S.~Deser and A.~Waldron, \emph{{Arbitrary spin representations in de Sitter
  from dS / CFT with applications to dS supergravity}},
  \href{https://doi.org/10.1016/S0550-3213(03)00348-1}{\emph{Nucl. Phys. B}
  {\bfseries 662} (2003) 379}
  [\href{https://arxiv.org/abs/hep-th/0301068}{{\ttfamily hep-th/0301068}}].

\bibitem{Anninos:2011ui}
D.~Anninos, T.~Hartman and A.~Strominger, \emph{{Higher Spin Realization of the
  dS/CFT Correspondence}},
  \href{https://doi.org/10.1088/1361-6382/34/1/015009}{\emph{Class. Quant.
  Grav.} {\bfseries 34} (2017) 015009}
  [\href{https://arxiv.org/abs/1108.5735}{{\ttfamily 1108.5735}}].

\bibitem{Anninos:2017eib}
D.~Anninos, F.~Denef, R.~Monten and Z.~Sun, \emph{{Higher Spin de Sitter
  Hilbert Space}}, \href{https://doi.org/10.1007/JHEP10(2019)071}{\emph{JHEP}
  {\bfseries 10} (2019) 071}
  [\href{https://arxiv.org/abs/1711.10037}{{\ttfamily 1711.10037}}].

\bibitem{Hertog:2019uhy}
T.~Hertog, G.~Tartaglino-Mazzucchelli and G.~Venken, \emph{{Spinors in
  Supersymmetric dS/CFT}},
  \href{https://doi.org/10.1007/JHEP10(2019)117}{\emph{JHEP} {\bfseries 10}
  (2019) 117} [\href{https://arxiv.org/abs/1905.01322}{{\ttfamily
  1905.01322}}].

\bibitem{Pethybridge:2024qci}
B.J.~Pethybridge, \emph{{Notes on complex $q=2$ SYK}},
  \href{https://arxiv.org/abs/2403.04673}{{\ttfamily 2403.04673}}.

\bibitem{Thomas:1941aa}
L.H.~Thomas, \emph{On {Unitary} {Representations} of the {Group} of {De}
  {Sitter} {Space}}, \href{https://doi.org/10.2307/1968990}{\emph{The Annals of
  Mathematics} {\bfseries 42} (1941) 113}.

\bibitem{Bargmann:1946me}
V.~Bargmann, \emph{{Irreducible unitary representations of the Lorentz group}},
  \href{https://doi.org/10.2307/1969129}{\emph{Annals Math.} {\bfseries 48}
  (1947) 568}.

\bibitem{Gelfand:1947aa}
I.M.~Gel'fand and M.A.~Na{\u{\i}}mark, \emph{Unitary representations of the
  {Lorentz} group}, {\emph{Izv. Akad. Nauk SSSR, Ser. Mat.} {\bfseries 11}
  (1947) 411}.

\bibitem{Harish-Chandra:1947aa}
Harish-Chandra, \emph{Infinite irreducible representations of the lorentz
  group}, {\emph{Proceedings of the Royal Society of London. Series A,
  Mathematical and Physical Sciences} {\bfseries 189} (1947) 372}.

\bibitem{Newton:1950aa}
T.D.~Newton, \emph{A {Note} on the {Representations} of the {De} {Sitter}
  {Group}}, \href{https://doi.org/10.2307/1969376}{\emph{The Annals of
  Mathematics} {\bfseries 51} (1950) 730}.

\bibitem{Dixmier:1961aa}
J.~Dixmier, \emph{Repr{\'e}sentations int{\'e}grables du groupe de {De}
  {Sitter}}, \href{https://doi.org/10.24033/bsmf.1558}{\emph{Bulletin de la
  Soci{\'e}t{\'e} Math{\'e}matique de France} {\bfseries 89} (1961) 9}.

\bibitem{Hirai:1962aa}
T.~Hirai, \emph{On irreducible representations of the {Lorentz} group of $n$-th
  order}, \href{https://doi.org/10.3792/pja/1195523378}{\emph{Proceedings of
  the Japan Academy, Series A, Mathematical Sciences} {\bfseries 38} (1962) }.

\bibitem{Takahashi:1963aa}
R.~Takahashi, \emph{Sur les repr{\'e}sentations unitaires des groupes de
  {Lorentz} g{\'e}n{\'e}ralis{\'e}s},
  \href{https://doi.org/10.24033/bsmf.1598}{\emph{Bulletin de la
  Soci{\'e}t{\'e} Math{\'e}matique de France} {\bfseries 91} (1963) 289}.

\bibitem{Ottoson:1968aa}
U.~Ottoson, \emph{A classification of the unitary irreducible representations
  of $so_0(n,1)$},
  \href{https://doi.org/10.1007/BF01645858}{\emph{Communications in
  Mathematical Physics} {\bfseries 8} (1968) 228}.

\bibitem{Gavrilik:1975aa}
A.M.~Gavrilik and A.U.~Klimyk, \emph{Analysis of the representations of the
  lorentz and euclidean groups of n-th order}, .

\bibitem{Dobrev:1977qv}
V.K.~Dobrev, G.~Mack, V.B.~Petkova, S.G.~Petrova and I.T.~Todorov,
  \emph{{Harmonic Analysis on the n-Dimensional Lorentz Group and Its
  Application to Conformal Quantum Field Theory}}, vol.~63, Springer (1977),
  \href{https://doi.org/10.1007/BFb0009678}{10.1007/BFb0009678}.

\bibitem{Sun:2021thf}
Z.~Sun, \emph{{A note on the representations of $\text{SO}(1,d+1)$}},
  \href{https://arxiv.org/abs/2111.04591}{{\ttfamily 2111.04591}}.

\bibitem{Basile:2017aa}
T.~Basile, X.~Bekaert and N.~Boulanger, \emph{Mixed-symmetry fields in de
  sitter space: a group theoretical glance},
  \href{https://doi.org/10.1007/JHEP05(2017)081}{\emph{JHEP} {\bfseries 2017}
  (2017) 81}.

\bibitem{Deser:2001us}
S.~Deser and A.~Waldron, \emph{{Partial masslessness of higher spins in
  (A)dS}}, \href{https://doi.org/10.1016/S0550-3213(01)00212-7}{\emph{Nucl.
  Phys. B} {\bfseries 607} (2001) 577}
  [\href{https://arxiv.org/abs/hep-th/0103198}{{\ttfamily hep-th/0103198}}].

\bibitem{Bonifacio:2023prb}
J.~Bonifacio and K.~Hinterbichler, \emph{{Fermionic Shift Symmetries in (Anti)
  de Sitter Space}},  \href{https://arxiv.org/abs/2312.06743}{{\ttfamily
  2312.06743}}.

\bibitem{Bonifacio:2018zex}
J.~Bonifacio, K.~Hinterbichler, A.~Joyce and R.A.~Rosen, \emph{{Shift
  Symmetries in (Anti) de Sitter Space}},
  \href{https://doi.org/10.1007/JHEP02(2019)178}{\emph{JHEP} {\bfseries 02}
  (2019) 178} [\href{https://arxiv.org/abs/1812.08167}{{\ttfamily
  1812.08167}}].

\bibitem{Bonifacio:2021mrf}
J.~Bonifacio, K.~Hinterbichler, A.~Joyce and D.~Roest, \emph{{Exceptional
  scalar theories in de Sitter space}},
  \href{https://doi.org/10.1007/JHEP04(2022)128}{\emph{JHEP} {\bfseries 04}
  (2022) 128} [\href{https://arxiv.org/abs/2112.12151}{{\ttfamily
  2112.12151}}].

\bibitem{Blauvelt:2022wwa}
E.~Blauvelt, L.~Engelbrecht and K.~Hinterbichler, \emph{{Shift Symmetries and
  AdS/CFT}}, \href{https://doi.org/10.1007/JHEP07(2023)103}{\emph{JHEP}
  {\bfseries 07} (2023) 103}
  [\href{https://arxiv.org/abs/2211.02055}{{\ttfamily 2211.02055}}].

\bibitem{Letsios:2023qzq}
V.A.~Letsios, \emph{{(Non-)unitarity of strictly and partially massless
  fermions on de Sitter space}},
  \href{https://doi.org/10.1007/JHEP05(2023)015}{\emph{JHEP} {\bfseries 05}
  (2023) 015} [\href{https://arxiv.org/abs/2303.00420}{{\ttfamily
  2303.00420}}].

\bibitem{Letsios:2023awz}
V.A.~Letsios, \emph{{New conformal-like symmetry of strictly massless fermions
  in four-dimensional de Sitter space}},
  \href{https://arxiv.org/abs/2310.01702}{{\ttfamily 2310.01702}}.

\bibitem{Letsios:2024nmf}
V.A.~Letsios, \emph{{(Non-)unitarity of strictly and partially massless
  fermions on de Sitter space II: an explanation based on the group-theoretic
  properties of the spin-3/2 and spin-5/2 eigenmodes}},
  \href{https://doi.org/10.1088/1751-8121/ad2c27}{\emph{J. Phys. A} {\bfseries
  57} (2024) 135401}.

\bibitem{Pilch:1984aw}
K.~Pilch, P.~van Nieuwenhuizen and M.F.~Sohnius, \emph{{De Sitter Superalgebras
  and Supergravity}}, \href{https://doi.org/10.1007/BF01211046}{\emph{Commun.
  Math. Phys.} {\bfseries 98} (1985) 105}.

\bibitem{Anous:2014lia}
T.~Anous, D.Z.~Freedman and A.~Maloney, \emph{{de Sitter Supersymmetry
  Revisited}}, \href{https://doi.org/10.1007/JHEP07(2014)119}{\emph{JHEP}
  {\bfseries 07} (2014) 119} [\href{https://arxiv.org/abs/1403.5038}{{\ttfamily
  1403.5038}}].

\bibitem{Bergshoeff:2015tra}
E.A.~Bergshoeff, D.Z.~Freedman, R.~Kallosh and A.~Van~Proeyen, \emph{{Pure de
  Sitter Supergravity}},
  \href{https://doi.org/10.1103/PhysRevD.93.069901}{\emph{Phys. Rev. D}
  {\bfseries 92} (2015) 085040}
  [\href{https://arxiv.org/abs/1507.08264}{{\ttfamily 1507.08264}}].

\bibitem{Anninos:2023exn}
D.~Anninos, P.~Benetti~Genolini and B.~M\"uhlmann, \emph{{dS$_{2}$
  supergravity}}, \href{https://doi.org/10.1007/JHEP11(2023)145}{\emph{JHEP}
  {\bfseries 11} (2023) 145}
  [\href{https://arxiv.org/abs/2309.02480}{{\ttfamily 2309.02480}}].

\bibitem{Dobrev:1976vr}
V.K.~Dobrev, G.~Mack, I.T.~Todorov, V.B.~Petkova and S.G.~Petrova, \emph{{On
  the Clebsch-Gordan Expansion for the Lorentz Group in $n$ Dimensions}},
  \href{https://doi.org/10.1016/0034-4877(76)90057-4}{\emph{Rept. Math. Phys.}
  {\bfseries 9} (1976) 219}.

\bibitem{Kravchuk:2016qvl}
P.~Kravchuk and D.~Simmons-Duffin, \emph{{Counting Conformal Correlators}},
  \href{https://doi.org/10.1007/JHEP02(2018)096}{\emph{JHEP} {\bfseries 02}
  (2018) 096} [\href{https://arxiv.org/abs/1612.08987}{{\ttfamily
  1612.08987}}].

\bibitem{Karateev:2018oml}
D.~Karateev, P.~Kravchuk and D.~Simmons-Duffin, \emph{{Harmonic Analysis and
  Mean Field Theory}},
  \href{https://doi.org/10.1007/JHEP10(2019)217}{\emph{JHEP} {\bfseries 10}
  (2019) 217} [\href{https://arxiv.org/abs/1809.05111}{{\ttfamily
  1809.05111}}].

\bibitem{Karateev:2018uk}
D.~Karateev, P.~Kravchuk and D.~Simmons-Duffin, \emph{Weight {Shifting}
  {Operators} and {Conformal} {Blocks}},
  \href{https://doi.org/10.1007/JHEP02(2018)081}{\emph{Journal of High Energy
  Physics} {\bfseries 2018} (2018) 81}.

\bibitem{Iliesiu:2015qra}
L.~Iliesiu, F.~Kos, D.~Poland, S.S.~Pufu, D.~Simmons-Duffin and R.~Yacoby,
  \emph{{Bootstrapping 3D Fermions}},
  \href{https://doi.org/10.1007/JHEP03(2016)120}{\emph{JHEP} {\bfseries 03}
  (2016) 120} [\href{https://arxiv.org/abs/1508.00012}{{\ttfamily
  1508.00012}}].

\bibitem{Iliesiu:2017nrv}
L.~Iliesiu, F.~Kos, D.~Poland, S.S.~Pufu and D.~Simmons-Duffin,
  \emph{{Bootstrapping 3D Fermions with Global Symmetries}},
  \href{https://doi.org/10.1007/JHEP01(2018)036}{\emph{JHEP} {\bfseries 01}
  (2018) 036} [\href{https://arxiv.org/abs/1705.03484}{{\ttfamily
  1705.03484}}].

\bibitem{Pethybridge:2021rwf}
B.~Pethybridge and V.~Schaub, \emph{{Tensors and spinors in de Sitter space}},
  \href{https://doi.org/10.1007/JHEP06(2022)123}{\emph{JHEP} {\bfseries 06}
  (2022) 123} [\href{https://arxiv.org/abs/2111.14899}{{\ttfamily
  2111.14899}}].

\bibitem{Gell-Mann:1961omu}
M.~Gell-Mann, \emph{{The Eightfold Way: A Theory of strong interaction
  symmetry}}, .

\bibitem{Anninos:2023lin}
D.~Anninos, T.~Anous, B.~Pethybridge and G.~\c{S}eng\"or, \emph{{The discreet
  charm of the discrete series in dS$_{2}$}},
  \href{https://doi.org/10.1088/1751-8121/ad14ad}{\emph{J. Phys. A} {\bfseries
  57} (2024) 025401} [\href{https://arxiv.org/abs/2307.15832}{{\ttfamily
  2307.15832}}].

\bibitem{Sengor:2023buj}
G.~\c{S}eng\"or, \emph{{Searching for discrete series representations at the
  late-time boundary of de Sitter}},  in \emph{{15th International Workshop on
  Lie Theory and Its Applications in Physics}}, 12, 2023
  [\href{https://arxiv.org/abs/2312.00363}{{\ttfamily 2312.00363}}].

\bibitem{RiosFukelman:2023mgq}
A.~Rios~Fukelman, M.~Semp\'e and G.A.~Silva, \emph{{Notes on gauge fields and
  discrete series representations in de Sitter spacetimes}},
  \href{https://doi.org/10.1007/JHEP01(2024)011}{\emph{JHEP} {\bfseries 01}
  (2024) 011} [\href{https://arxiv.org/abs/2310.14955}{{\ttfamily
  2310.14955}}].

\bibitem{Hirai:1965aa}
T.~Hirai, \emph{The characters of irreducible representations of the {Lorentz}
  group of $n$-th order},
  \href{https://doi.org/10.3792/pja/1195522333}{\emph{Proceedings of the Japan
  Academy, Series A, Mathematical Sciences} {\bfseries 41} (1965) }.

\bibitem{Penedones:2023uqc}
J.~Penedones, K.~Salehi~Vaziri and Z.~Sun, \emph{{Hilbert space of Quantum
  Field Theory in de Sitter spacetime}},
  \href{https://arxiv.org/abs/2301.04146}{{\ttfamily 2301.04146}}.

\bibitem{Kitaev:2017hnr}
A.~Kitaev, \emph{{Notes on $\widetilde{\mathrm{SL}}(2,\mathbb{R})$
  representations}},  \href{https://arxiv.org/abs/1711.08169}{{\ttfamily
  1711.08169}}.

\bibitem{Kravchuk:2021akc}
P.~Kravchuk, D.~Mazac and S.~Pal, \emph{{Automorphic Spectra and the Conformal
  Bootstrap}}, \href{https://doi.org/10.1090/cams/26}{\emph{Commun. Am. Math.
  Soc.} {\bfseries 4} (2024) 1}
  [\href{https://arxiv.org/abs/2111.12716}{{\ttfamily 2111.12716}}].

\bibitem{Knapp:1971aa}
A.W.~Knapp and E.M.~Stein, \emph{Interwining {Operators} for {Semisimple}
  {Groups}}, \href{https://doi.org/10.2307/1970887}{\emph{The Annals of
  Mathematics} {\bfseries 93} (1971) 489}.

\bibitem{Knapp:1982aa}
A.W.~Knapp and G.J.~Zuckerman, \emph{Classification of {Irreducible} {Tempered}
  {Representations} of {Semisimple} {Groups}},
  \href{https://doi.org/10.2307/2007019}{\emph{The Annals of Mathematics}
  {\bfseries 116} (1982) 457}.

\bibitem{Knapp:1986aa}
A.W.~Knapp, \emph{Representation Theory of Semisimple Groups: An Overview Based
  on Examples (PMS-36)}, Princeton University Press, rev - revised~ed. (1986).

\bibitem{Casselman:1989aa}
W.~Casselman, \emph{Canonical extensions of harish-chandra modules to
  representations of g}, {\emph{Canadian Journal of Mathematics} {\bfseries 41}
  (1989) 385}.

\bibitem{Ferrara:1972kab}
S.~Ferrara, A.F.~Grillo, G.~Parisi and R.~Gatto, \emph{{Covariant expansion of
  the conformal four-point function}},
  \href{https://doi.org/10.1016/0550-3213(73)90467-7}{\emph{Nucl. Phys. B}
  {\bfseries 49} (1972) 77}.

\bibitem{Dolan:2004up}
F.A.~Dolan and H.~Osborn, \emph{Conformal {Partial} {Waves} and the {Operator}
  {Product} {Expansion}},
  \href{https://doi.org/10.1016/j.nuclphysb.2003.11.016}{\emph{Nuclear Physics
  B} {\bfseries 678} (2004) 491}.

\bibitem{Simmons-Duffin:2014wb}
D.~Simmons-Duffin, \emph{Projectors, {Shadows}, and {Conformal} {Blocks}},
  \href{https://doi.org/10.1007/JHEP04(2014)146}{\emph{Journal of High Energy
  Physics} {\bfseries 2014} (2014) 146}.

\bibitem{Costa:2011wa}
M.S.~Costa, J.~Penedones, D.~Poland and S.~Rychkov, \emph{Spinning {Conformal}
  {Correlators}}, \href{https://doi.org/10.1007/JHEP11(2011)071}{\emph{Journal
  of High Energy Physics} {\bfseries 2011} (2011) 71}.

\bibitem{Bargmann:1977gy}
V.~Bargmann and I.T.~Todorov, \emph{{Spaces of Analytic Functions on a Complex
  Cone as Carries for the Symmetric Tensor Representations of SO(N)}},
  \href{https://doi.org/10.1063/1.523383}{\emph{J. Math. Phys.} {\bfseries 18}
  (1977) 1141}.

\bibitem{Dolan:2001wg}
F.A.~Dolan and H.~Osborn, \emph{Conformal {Four} {Point} {Functions} and the
  {Operator} {Product} {Expansion}},
  \href{https://doi.org/10.1016/S0550-3213(01)00013-X}{\emph{Nuclear Physics B}
  {\bfseries 599} (2001) 459}.

\bibitem{Dolan:2012wt}
F.A.~Dolan and H.~Osborn, \emph{Conformal {Partial} {Waves}: {Further}
  {Mathematical} {Results}}, {\emph{arXiv:1108.6194 [hep-th]} (2012) }.

\bibitem{Costa:2014rya}
M.S.~Costa and T.~Hansen, \emph{{Conformal correlators of mixed-symmetry
  tensors}}, \href{https://doi.org/10.1007/JHEP02(2015)151}{\emph{JHEP}
  {\bfseries 02} (2015) 151} [\href{https://arxiv.org/abs/1411.7351}{{\ttfamily
  1411.7351}}].

\bibitem{Gradshtein:2015aa}
I.S.~Gradshte{\u \i}n and D.~Zwillinger, \emph{Table of integrals, series, and
  products}, Elsevier, Academic Press is an imprint of Elsevier, Amsterdam ;
  Boston, eighth edition~ed. (2015).

\bibitem{Weinberg:2010ws}
S.~Weinberg, \emph{Six-dimensional {Methods} for {Four}-dimensional {Conformal}
  {Field} {Theories}},
  \href{https://doi.org/10.1103/PhysRevD.82.045031}{\emph{Physical Review D}
  {\bfseries 82} (2010) 045031}.

\bibitem{Bzowski:2013sza}
A.~Bzowski, P.~McFadden and K.~Skenderis, \emph{{Implications of conformal
  invariance in momentum space}},
  \href{https://doi.org/10.1007/JHEP03(2014)111}{\emph{JHEP} {\bfseries 03}
  (2014) 111} [\href{https://arxiv.org/abs/1304.7760}{{\ttfamily 1304.7760}}].

\bibitem{Bzowski:2015pba}
A.~Bzowski, P.~McFadden and K.~Skenderis, \emph{{Scalar 3-point functions in
  CFT: renormalisation, beta functions and anomalies}},
  \href{https://doi.org/10.1007/JHEP03(2016)066}{\emph{JHEP} {\bfseries 03}
  (2016) 066} [\href{https://arxiv.org/abs/1510.08442}{{\ttfamily
  1510.08442}}].

\bibitem{Bzowski:2018fql}
A.~Bzowski, P.~McFadden and K.~Skenderis, \emph{{Renormalised CFT 3-point
  functions of scalars, currents and stress tensors}},
  \href{https://doi.org/10.1007/JHEP11(2018)159}{\emph{JHEP} {\bfseries 11}
  (2018) 159} [\href{https://arxiv.org/abs/1805.12100}{{\ttfamily
  1805.12100}}].

\bibitem{Costa:2011dw}
M.S.~Costa, J.~Penedones, D.~Poland and S.~Rychkov, \emph{{Spinning Conformal
  Blocks}}, \href{https://doi.org/10.1007/JHEP11(2011)154}{\emph{JHEP}
  {\bfseries 11} (2011) 154} [\href{https://arxiv.org/abs/1109.6321}{{\ttfamily
  1109.6321}}].

\bibitem{Costa:2018mcg}
M.S.~Costa and T.~Hansen, \emph{{AdS Weight Shifting Operators}},
  \href{https://doi.org/10.1007/JHEP09(2018)040}{\emph{JHEP} {\bfseries 09}
  (2018) 040} [\href{https://arxiv.org/abs/1805.01492}{{\ttfamily
  1805.01492}}].

\bibitem{Bros:2010wa}
J.~Bros, H.~Epstein and U.~Moschella, \emph{{Scalar tachyons in the de Sitter
  universe}}, \href{https://doi.org/10.1007/s11005-010-0406-4}{\emph{Lett.
  Math. Phys.} {\bfseries 93} (2010) 203}
  [\href{https://arxiv.org/abs/1003.1396}{{\ttfamily 1003.1396}}].

\bibitem{Epstein:2014jaa}
H.~Epstein and U.~Moschella, \emph{{de Sitter tachyons and related topics}},
  \href{https://doi.org/10.1007/s00220-015-2308-x}{\emph{Commun. Math. Phys.}
  {\bfseries 336} (2015) 381}
  [\href{https://arxiv.org/abs/1403.3319}{{\ttfamily 1403.3319}}].

\bibitem{Isono:2017grm}
H.~Isono, \emph{{On conformal correlators and blocks with spinors in general
  dimensions}}, \href{https://doi.org/10.1103/PhysRevD.96.065011}{\emph{Phys.
  Rev. D} {\bfseries 96} (2017) 065011}
  [\href{https://arxiv.org/abs/1706.02835}{{\ttfamily 1706.02835}}].

\bibitem{Clif:Van-Lancker:2001aa}
P.~Van~Lancker, F.~Sommen and D.~Constales, \emph{Models for irreducible
  representations of ${Spin}(m)$},
  \href{https://doi.org/10.1007/BF03042223}{\emph{Advances in Applied Clifford
  Algebras} {\bfseries 11} (2001) 271}.

\bibitem{Clif1}
V.~Dietrich, K.~Habetha and G.~Jank, eds., \emph{Clifford {Algebras} and
  {Their} {Application} in {Mathematical} {Physics}: {Aachen} 1996}, Springer
  Netherlands, Dordrecht (1998),
  \href{https://doi.org/10.1007/978-94-011-5036-1}{10.1007/978-94-011-5036-1}.

\bibitem{Clif2}
J.~Ryan and W.~Spr{\"o}{\ss}ig, eds., \emph{Clifford {Algebras} and their
  {Applications} in {Mathematical} {Physics}}, Birkh{\"a}user Boston, Boston,
  MA (2000),
  \href{https://doi.org/10.1007/978-1-4612-1374-1}{10.1007/978-1-4612-1374-1}.

\bibitem{Clif:Delanghe:1992aa}
R.~Delanghe, F.~Sommen and V.~Sou{\v c}ek, \emph{Clifford {Algebra} and
  {Spinor}-{Valued} {Functions}}, Springer Netherlands, Dordrecht (1992),
  \href{https://doi.org/10.1007/978-94-011-2922-0}{10.1007/978-94-011-2922-0}.

\bibitem{Clif:Van-Lancker:2009aa}
P.~Van~Lancker, \emph{Spherical {Monogenics}: {An} {Algebraic} {Approach}},
  \href{https://doi.org/10.1007/s00006-009-0168-1}{\emph{Advances in Applied
  Clifford Algebras} {\bfseries 19} (2009) 467}.

\bibitem{Clif:Van-Lancker:2015aa}
P.~Van~Lancker, \emph{Spinor-valued $l^2$-spaces on $s_{m-1}$ and
  representations of $spin^{+}(1,m)$},
  \href{https://doi.org/10.1016/j.jmaa.2014.09.066}{\emph{Journal of
  Mathematical Analysis and Applications} {\bfseries 423} (2015) 253}.

\bibitem{Erramilli:2019njx}
R.S.~Erramilli, L.V.~Iliesiu and P.~Kravchuk, \emph{{Recursion relation for
  general 3d blocks}},
  \href{https://doi.org/10.1007/JHEP12(2019)116}{\emph{JHEP} {\bfseries 12}
  (2019) 116} [\href{https://arxiv.org/abs/1907.11247}{{\ttfamily
  1907.11247}}].

\bibitem{Duc:1967aa}
D.V.~Duc and N.V.~Hieu, \emph{On the theory of unitary representations of the
  ${SL}(2, {C})$ group}, {\emph{Annales de l'institut Henri Poincar{\'e}.
  Section A, Physique Th{\'e}orique} {\bfseries 6} (1967) 17}.

\bibitem{Bonifacio:2023ban}
J.~Bonifacio, D.~Mazac and S.~Pal, \emph{{Spectral Bounds on Hyperbolic
  3-Manifolds: Associativity and the Trace Formula}},
  \href{https://arxiv.org/abs/2308.11174}{{\ttfamily 2308.11174}}.

\bibitem{DiFrancesco:1997nk}
P.~Di~Francesco, P.~Mathieu and D.~Senechal, \emph{{Conformal Field Theory}},
  Graduate Texts in Contemporary Physics, Springer-Verlag, New York (1997),
  \href{https://doi.org/10.1007/978-1-4612-2256-9}{10.1007/978-1-4612-2256-9}.

\bibitem{Gerry_Knight_2004}
C.~Gerry and P.~Knight, \emph{Introductory Quantum Optics}, Cambridge
  University Press (2004).

\bibitem{Weinberg:1980kq}
S.~Weinberg and E.~Witten, \emph{{Limits on Massless Particles}},
  \href{https://doi.org/10.1016/0370-2693(80)90212-9}{\emph{Phys. Lett. B}
  {\bfseries 96} (1980) 59}.

\bibitem{Harish-Chandra:1966aa}
Harish-Chandra, \emph{{Discrete series for semisimple Lie groups. II: Explicit
  determination of the characters}},
  \href{https://doi.org/10.1007/BF02392813}{\emph{Acta Mathematica} {\bfseries
  116} (1966) 1 }.

\bibitem{Du:2004jt}
D.-P.~Du, B.~Wang and R.-K.~Su, \emph{{Quasinormal modes in pure de Sitter
  space-times}}, \href{https://doi.org/10.1103/PhysRevD.70.064024}{\emph{Phys.
  Rev. D} {\bfseries 70} (2004) 064024}
  [\href{https://arxiv.org/abs/hep-th/0404047}{{\ttfamily hep-th/0404047}}].

\bibitem{Lopez-Ortega:2006aal}
A.~Lopez-Ortega, \emph{{Quasinormal modes of D-dimensional de Sitter
  spacetime}}, \href{https://doi.org/10.1007/s10714-006-0335-9}{\emph{Gen. Rel.
  Grav.} {\bfseries 38} (2006) 1565}
  [\href{https://arxiv.org/abs/gr-qc/0605027}{{\ttfamily gr-qc/0605027}}].

\bibitem{Lopez-Ortega:2012xvr}
A.~Lopez-Ortega, \emph{{On the quasinormal modes of the de Sitter spacetime}},
  \href{https://doi.org/10.1007/s10714-012-1398-4}{\emph{Gen. Rel. Grav.}
  {\bfseries 44} (2012) 2387}
  [\href{https://arxiv.org/abs/1207.6791}{{\ttfamily 1207.6791}}].

\bibitem{Jafferis:2013qia}
D.L.~Jafferis, A.~Lupsasca, V.~Lysov, G.S.~Ng and A.~Strominger,
  \emph{{Quasinormal quantization in de Sitter spacetime}},
  \href{https://doi.org/10.1007/JHEP01(2015)004}{\emph{JHEP} {\bfseries 01}
  (2015) 004} [\href{https://arxiv.org/abs/1305.5523}{{\ttfamily 1305.5523}}].

\bibitem{Sun:2020sgn}
Z.~Sun, \emph{{Higher spin de Sitter quasinormal modes}},
  \href{https://doi.org/10.1007/JHEP11(2021)025}{\emph{JHEP} {\bfseries 11}
  (2021) 025} [\href{https://arxiv.org/abs/2010.09684}{{\ttfamily
  2010.09684}}].

\bibitem{Law:2023ohq}
Y.T.A.~Law, \emph{{Characters, Quasinormal Modes, and Quantum de Sitter
  Thermodynamics}}, \href{https://doi.org/10.22323/1.436.0130}{\emph{PoS}
  {\bfseries CORFU2022} (2023) 130}
  [\href{https://arxiv.org/abs/2304.01471}{{\ttfamily 2304.01471}}].

\bibitem{Anninos:2020hfj}
D.~Anninos, F.~Denef, Y.T.A.~Law and Z.~Sun, \emph{{Quantum de Sitter horizon
  entropy from quasicanonical bulk, edge, sphere and topological string
  partition functions}},
  \href{https://doi.org/10.1007/JHEP01(2022)088}{\emph{JHEP} {\bfseries 01}
  (2022) 088} [\href{https://arxiv.org/abs/2009.12464}{{\ttfamily
  2009.12464}}].

\bibitem{Law:2020cpj}
Y.T.A.~Law, \emph{{A compendium of sphere path integrals}},
  \href{https://doi.org/10.1007/JHEP12(2021)213}{\emph{JHEP} {\bfseries 21}
  (2020) 213} [\href{https://arxiv.org/abs/2012.06345}{{\ttfamily
  2012.06345}}].

\bibitem{Muhlmann:2021aa}
B.~M{\"u}hlmann, \emph{The two-sphere partition function in two-dimensional
  quantum gravity at fixed area},
  \href{https://doi.org/10.1007/JHEP09(2021)189}{\emph{JHEP} {\bfseries 09}
  (2021) }.

\bibitem{Anninos:2019aa}
D.~Anninos, D.A.~Galante and D.M.~Hofman, \emph{De sitter horizons \&
  holographic liquids},
  \href{https://doi.org/10.1007/jhep07(2019)038}{\emph{Journal of High Energy
  Physics} {\bfseries 2019} (2019) }.

\bibitem{Albrychiewicz:2020ruh}
E.~Albrychiewicz and Y.~Neiman, \emph{{Scattering in the static patch of de
  Sitter space}},
  \href{https://doi.org/10.1103/PhysRevD.103.065014}{\emph{Phys. Rev. D}
  {\bfseries 103} (2021) 065014}
  [\href{https://arxiv.org/abs/2012.13584}{{\ttfamily 2012.13584}}].

\bibitem{Harris2021}
D.~Anninos and E.~Harris, \emph{Three-dimensional de sitter horizon
  thermodynamics}, \href{https://doi.org/10.1007/JHEP10(2021)091}{\emph{Journal
  of High Energy Physics 2021 2021:10} {\bfseries 2021} (2021) 1}.

\bibitem{Mirbabayi:2022gnl}
M.~Mirbabayi and F.~Riccardi, \emph{{Probing de Sitter from the horizon}},
  \href{https://arxiv.org/abs/2211.11672}{{\ttfamily 2211.11672}}.

\bibitem{Harish-Chandra:1965aa}
Harish-Chandra, \emph{Discrete series for semisimple {Lie} groups. {I}:
  {Construction} of invariant eigendistributions},
  \href{https://doi.org/10.1007/BF02391779}{\emph{Acta Math.} {\bfseries 113}
  (1965) 241}.

\end{thebibliography}\endgroup

\end{document}